\documentclass[twocolumn]{aastex6}

\usepackage{comment}
\usepackage{amssymb}
\usepackage{graphicx}
\usepackage{multirow}
\usepackage{subcaption}
\usepackage{tikz}
\tikzset{>=latex}
\usetikzlibrary{arrows,calc,positioning}

\newcommand{\about}{$\sim$ }
\newcommand{\amin}{$^{\prime}$ }

\newcommand{\bces}{{\tt BCES}}
\newcommand{\chandra}{\emph{Chandra}}

\newcommand{\mekal}{{\tt MeKaL}}
\newcommand{\mfive}{$M_{500}$}

\newcommand{\rfive}{$R_{500}$}

\newcommand{\xmm}{\emph{XMM-Newton}}

\newcommand{\xray}{X-ray}
\newcommand{\xspec}{{\tt XSPEC}}

\shorttitle{XMM-DLS: X-ray vs. Weak Lensing Masses}
\shortauthors{Deshpande et al.}

\defcitealias{wittman06}{{\tt W}06}
\defcitealias{hilton12}{H12}
\defcitealias{maugh12}{M12}
\defcitealias{abate09}{A09}

\begin{document}

%%%%%%%%%%%%%%%%%%%%%%%%%%%%%%%%%%%%%%%%%%%%%%%%%%%%%%%%%%%%%%%%%%%%%%%
%%%%%%%%%%%%%%%%%%%%%%%%%%%%%%%%%%%%%%%%%%%%%%%%%%%%%%%%%%%%%%%%%%%%%%%
% %
% %		TITLE 
% %
%%%%%%%%%%%%%%%%%%%%%%%%%%%%%%%%%%%%%%%%%%%%%%%%%%%%%%%%%%%%%%%%%%%%%%%
%%%%%%%%%%%%%%%%%%%%%%%%%%%%%%%%%%%%%%%%%%%%%%%%%%%%%%%%%%%%%%%%%%%%%%%

\title{X-ray Temperatures, Luminosities, and Masses from \xmm\ Follow-up of the first shear-selected galaxy cluster sample\footnotemark[1]}

\footnotetext[1]{Some of the data presented herein were obtained at the W.M. Keck Observatory, which is operated as a scientific partnership among the California Institute of Technology, the University of California and  the National Aeronautics and Space Administration. The Observatory was made possible by the generous financial support of the W.M. Keck Foundation.\\}

%%%%%%%%%%%%%%%%%%%%%%%%%%%%%%%%%%%%%%%%%%%%%%%%%%%%%%%%%%%%%%%%%%%%%%%
%%%%%%%%%%%%%%%%%%%%%%%%%%%%%%%%%%%%%%%%%%%%%%%%%%%%%%%%%%%%%%%%%%%%%%%
% %
% %		AUTHORS AND AFFILS
% %
%%%%%%%%%%%%%%%%%%%%%%%%%%%%%%%%%%%%%%%%%%%%%%%%%%%%%%%%%%%%%%%%%%%%%%%
%%%%%%%%%%%%%%%%%%%%%%%%%%%%%%%%%%%%%%%%%%%%%%%%%%%%%%%%%%%%%%%%%%%%%%%

\author{Amruta~J.~Deshpande\altaffilmark{2}, John~P.~Hughes\altaffilmark{3}}
\affil{Department of Physics and Astronomy, Rutgers the State
        University of New Jersey, Piscataway, NJ 08904}
\altaffiltext{2}{amrejd@physics.rutgers.edu}
\altaffiltext{3}{jph@physics.rutgers.edu}
        \and
        \author{David Wittman\altaffilmark{4}$^,$\altaffilmark{5}}
\affil{Department of Physics, University of California, Davis, One Shields Ave., Davis, CA 95616}
\altaffiltext{4}{dwittman@physics.ucdavis.edu}
\altaffiltext{5}{also,
Instituto de  Astrof{\' i}sica e Ci{\^ e}ncias de Espa{\c c}o, Universidade de Lisboa}

%%%%%%%%%%%%%%%%%%%%%%%%%%%%%%%%%%%%%%%%%%%%%%%%%%%%%%%%%%%%%%%%%%%%%%%
%%%%%%%%%%%%%%%%%%%%%%%%%%%%%%%%%%%%%%%%%%%%%%%%%%%%%%%%%%%%%%%%%%%%%%%
% %
% %		ABSTRACT 
% %
%%%%%%%%%%%%%%%%%%%%%%%%%%%%%%%%%%%%%%%%%%%%%%%%%%%%%%%%%%%%%%%%%%%%%%%
%%%%%%%%%%%%%%%%%%%%%%%%%%%%%%%%%%%%%%%%%%%%%%%%%%%%%%%%%%%%%%%%%%%%%%%

\begin{abstract}
We continue the study of the first sample of shear-selected clusters \citep{wittman06} from the initial $8.6$~square degrees of the Deep Lens Survey \citep[DLS,][]{wittman02}; a sample with well-defined selection criteria corresponding to the highest ranked shear peaks in the survey area. 
We aim to characterize the weak lensing selection by examining the sample's \xray\ properties.
There are multiple X-ray clusters associated with nearly all the shear peaks: $14$ X-ray clusters corresponding to seven DLS shear peaks.
An additional three X-ray clusters cannot be definitively associated with shear peaks, mainly due to large positional offsets between the X-ray centroid and the shear peak. 
Here we report on the \xmm\ properties of the $17$ X-ray clusters. 
The X-ray clusters display a wide range of luminosities and temperatures; the $L_X-T_X$ relation we determine for the shear-associated X-ray clusters is consistent with X-ray cluster samples selected without regard to dynamical state, while it is inconsistent with self-similarity. 
For a subset of the sample, we measure \xray\ masses using temperature as a proxy, and compare to weak lensing masses determined by the DLS team  \citep{abate09,witt14}. 
The resulting mass comparison is consistent with equality.
The X-ray and weak lensing masses show considerable intrinsic scatter ($\sim48\%$), which is consistent with X-ray selected samples when their X-ray and weak lensing masses are independently determined. 

\end{abstract}

%%%%%%%%%%%%%%%%%%%%%%%%%%%%%%%%%%%%%%%%%%%%%%%%%%%%%%%%%%%%%%%%%%%%%%%
%%%%%%%%%%%%%%%%%%%%%%%%%%%%%%%%%%%%%%%%%%%%%%%%%%%%%%%%%%%%%%%%%%%%%%%
% %
% %		INTRODUCTION 
% %
%%%%%%%%%%%%%%%%%%%%%%%%%%%%%%%%%%%%%%%%%%%%%%%%%%%%%%%%%%%%%%%%%%%%%%%
%%%%%%%%%%%%%%%%%%%%%%%%%%%%%%%%%%%%%%%%%%%%%%%%%%%%%%%%%%%%%%%%%%%%%%%

\section{Introduction}

Clusters of galaxies are a sensitive probe of cosmology \citep[e.g., ][]{allen11}.
They have been widely observed through the baryonic signatures of their galaxies or intracluster gas. 
The cosmologically relevant quantity, cluster mass, is dominated by a non-baryonic component, dark matter, at a ratio of approximately $5\:$:$1$.
The total cluster mass, when inferred from baryonic components, is limited in accuracy by required assumptions (e.g., hydrostatic equilibrium for the gas and virial equilibrium for the galaxy velocity distribution).  
Typically, small, well-studied cluster samples are used to calibrate scaling laws for the total mass and specific mass observables (e.g., gas temperature, velocity dispersion) that can then be applied to less well-studied systems.
Total cluster mass can also be measured using the statistical distortion of background galaxies by gravitational lensing  (weak lensing), which is  directly sensitive to the mass without a dependence on the properties of the baryonic cluster components.  
Understanding the relationship between multiple mass proxies to arrive at a better estimate for cluster mass is required in order to use clusters to constrain the growth of structure in the Universe.

The power of finding clusters directly through the fundamental cosmological quantity, the cluster mass, was recognized by early weak lensing studies \citep{tyson90,kaiser92}.  
Weak lensing selects solely on the projected mass along a line of sight and is largely independent of physical processes that can affect our observations of the baryonic components (e.g., mergers).  
A large number of individual clusters have been studied in shear, but there have been fewer studies of shear-{\it selected} clusters \citep{wittman06,miyazaki07,gavsouc07,schirmer07,miyazaki15}.   
The first set of clusters selected in shear was published by \citet{wittman06} from the Deep Lens Survey \citep[hereafter DLS;][]{wittman02}.  
 
Although there have been numerous weak lensing follow-up studies of X-ray or optically selected samples, follow-up efforts that focus on characterizing the properties of weak lensing selected clusters are few in the literature \citep[e.g.,][9 clusters]{giles15}. 
Our work with the DLS falls in this latter camp.

We continue the study of the shear-selected clusters discovered in \citet{wittman06}. 
These are 7 of the 8 highest ranked shear peaks in the first $8.6$~deg$^2$ of the $20$~deg$^2$ DLS.
The top ranked shear peak among them corresponds to the previously known complex of clusters associated with Abell~781.
This complex has been previously studied in detail, both in \xray s and in weak lensing with emphasis on mass comparison \citep{sehg08,witt14}.
The fifth ranked shear peak was deemed to be a line of sight projection, while the remaining $6$ have all been confirmed as clusters.
The majority of the shear peaks show multiple \xray\ and optical (in the DLS) counterparts \citep{wittman06}.

The initial follow-up to confirm the shear peaks as  clusters was conducted by \citet{wittman06}, using low exposure \chandra\ imaging.
We have since been awarded \xmm\ data, with which we can learn more by examining the sample in some of the best studied (and low scatter) \xray\ properties: $L_X$, the \xray\ luminosity, and $T_X$, the \xray\ temperature \citep[e.g.,][]{vikh09,pratt09,mantz10}.  
We can examine them as mass proxies \citep{ettori13} and study their behavior along \xray\ scaling laws \citep[e.g.,][]{pratt09,maugh12,mahdavi13}, which are typically low in scatter and drawn from self-generated properties.

In this study we determine \xray\ temperatures, luminosities, and masses.
Our sample covers the same survey area as \citet{wittman06}, hereafter \citetalias{wittman06}, but goes further into the distribution of shear, adding three more peaks.
Some of the DLS fields in our study (in particular F2) have previously been examined, in part or in entirety, by other studies \citep{kubo09,utsumi14,miyazaki15,geller10,starik14,ascaso14}; we discuss them in the context of our own work in section \S\ref{subsec-SourceDetect} below. 
Our study includes DLS fields F2-F5, encompassing a larger survey area than these other studies. 
We focus on the X-ray properties of the sample, showing the $L_X-T_X$ relation for the first time and comparing it to X-ray selected cluster samples. 
We obtain X-ray mass estimates using temperature as a proxy which we compare to weak lensing masses determined by the DLS team \citep{abate09,witt14}.

This paper is organized as follows.  
The \xray\ data, its analysis and the cluster properties are discussed in section \ref{sec-xmmdata}.  
The luminosity-temperature relation is presented in \S\ref{subsec-lxtx}. 
The \xray\ mass estimates and comparison to weak lensing are discussed in section~\ref{sec-mass}.  
We conclude with a summary in \S\ref{sec-summ}.  
Throughout this paper we use  $H_0$ = 70 km~s$^{-1}$Mpc$^{-1}$, $h=H_0/(100 \textnormal{ km s}^{-1} \textnormal{Mpc}^{-1})$, $\Omega_{\Lambda} = 0.7$ and $\Omega_{m,0} = 0.3$, $E(z)=\sqrt{(\Omega_{m,0}(1+z)^3 + \Omega_\Lambda)}$, and report all uncertainties at the $1\sigma$ confidence level.

%%%%%%%%%%%%%%%%%%%%%%%%%%%%%%%%%%%%%%%%%%%%%%%%%%%%%%%%%%%%%%%%%%%%%%%
%%%%%%%%%%%%%%%%%%%%%%%%%%%%%%%%%%%%%%%%%%%%%%%%%%%%%%%%%%%%%%%%%%%%%%%
% %
% %		SECTION - XMM data and analysis
% %
%%%%%%%%%%%%%%%%%%%%%%%%%%%%%%%%%%%%%%%%%%%%%%%%%%%%%%%%%%%%%%%%%%%%%%%
%%%%%%%%%%%%%%%%%%%%%%%%%%%%%%%%%%%%%%%%%%%%%%%%%%%%%%%%%%%%%%%%%%%%%%%

\section{ The Sample, X-RAY Data, \& Analysis} %
\label{sec-xmmdata}%

%%%%%%%%%%%%%%%%%%%%%%%%%%%%%%%%%
%	TABLE : obs_info
%%%%%%%%%%%%%%%%%%%%%%%%%%%%%%%%%

\begin{table}
\caption{\emph{XMM-Newton} observations.}
\resizebox{0.82\textwidth}{!}{\begin{minipage}{\textwidth}
\begin{tabular}{>{\hspace{-2ex}} c <{\hspace{-1.7ex}} l  c <{\hspace{-1.8ex}}  c <{\hspace{-3ex}} r  c <{\hspace{-3ex}}  r }
\tableline\\[-3.5ex] 
\tableline\\[-1.7ex]
No.				& Name  	& OBS\_IDS\hspace{4ex} 		& \multicolumn{2}{c}{Duration}			& \multicolumn{2}{c}{Exposure} 			\\
   				&	 		& 			& \multicolumn{2}{c}{(s)}			& \multicolumn{2}{c}{(s)}			\\ 
				&	 		&			&  PN		& $\left<\textup{MOS}\right>$	& PN		& $\left<\textup{MOS}\right>$	\\[2pt] 
\tableline\\[-2ex]
\multirow{2}{*}[0.5em]{1.} & \multirow{2}{*}[0.5em]{DLSCL~J0920.1+3029}	& 0150620201$^{\left(a\right)}$					& $13230$	& $16173$	& $11709$	& $14466$	\\ 
				& 		     											& 0401170101\phm{$^{\left(a\right)}$} 			& $68695$	& $78230$	& $52107$	& $67687$	\\
2.				& DLSCL~J0522.2$-$4820									& 0303820101\phm{$^{\left(a\right)}$}			& $34700$	& $41572$	& $8943$	& $24265$	\\ 
3.				& DLSCL~J1049.6$-$0417									& $^{\left(b\right)}$\phm{$^{\left(a\right)}$}	& \nodata  	& \nodata 	& \nodata 	& \nodata   \\
4.				& DLSCL~J1054.1$-$0549 									& 0552860101\phm{$^{\left(a\right)}$} 			& $51128$	& $53208$	& $29092$	& $36667$	\\ 
5. 				& DLSCL~J1402.2$-$1028									& $^{\left(b\right)}$\phm{$^{\left(a\right)}$}	& \nodata 	& \nodata  	& \nodata  	& \nodata   \\
6.				& DLSCL~J1402.0$-$1019									& $^{\left(b\right)}$\phm{$^{\left(a\right)}$} 	& \nodata 	& \nodata 	& \nodata  	& \nodata  	\\
7.				& DLSCL~J0916.0+2931 									& 0303820301\phm{$^{\left(a\right)}$} 			& $39937$	& $41572$	& $16982$	& $23393$	\\ 
8.				& DLSCL~J1055.2$-$0503 									& 0303820201\phm{$^{\left(a\right)}$}			& $34933$	& $36572$	& $28309$ 	& $30876$	\\

	 									\multicolumn{3}{r}{Averages:- }													& $40437$ 	& $44544$ 	& $24254$ 	& $32892$  	\\[2pt]   

\multicolumn{7}{l}{\emph{Beyond the initial \citet{wittman06} publication}: } \\[2pt]

$\;\;\;$B$9.\;\;$ 	& DLSCL~J1048.5$-$0411 								& 0150620901\phm{$^{\left(a\right)}$} 			& $12036$	& $13672$	& $10167$	& $12310$	\\ 
$\;\;\;$B$10.$      & DLSCL~J0921.4+3013								& 0150620101\phm{$^{\left(a\right)}$}			& $11268$	& $15781$	& $8936$	& $13000$	\\
$\;\;\;$B$11.$      & DLSCL~J0916.3+3025								& 0152060301\phm{$^{\left(a\right)}$} 			& $11605$   & $10239$   & $8605$	& $9142$    \\
 
 										\multicolumn{3}{r}{Averages:- }               									& $11636$ 	& $13231$ 	& $9236$ 	& $11484$ 	\\[2pt]%
\tableline
\end{tabular}
\end{minipage}
} % ends resizebox
\tablecomments{% 
Column (1) gives the DLS candidate number from \citetalias{wittman06}, or designations beginning with the letter B that we assign here to the \emph{beyond} subset. 
`Duration' reports the total telescope on-time.  
`Exposure' shows the total exposure after background flare filtering. 
`$\left<\textup{MOS}\right>$' gives the average value from the two mos cameras. 
An $\left(a\right)$ indicates the observation is analyzed in \citet{sehg08}.
A $\left(b\right)$ indicates there is no corresponding \xmm\ data; initial \emph{Chandra} follow-up \citepalias{wittman06} found no X-ray counterpart to peak 5, and found very low signal-to-noise X-rays corresponding to shear peaks 3 and 6.
}%
\label{tab-obs_info}%
\end{table}

The DLS is described in more detail in W06. Briefly, it is a 20 deg$^2$ $BVRz^\prime$ survey that is 50\% complete at $R=25.8$ (Vega), its deepest band. This places it intermediate to the CFHTLS-Wide\footnote[6]{Canada-France-Hawaii Telescope Legacy Survey's Deep and Wide fields: http://www.cfht.hawaii.edu/Science/CFHLS/ \\cfhtlsdeepwidefields.html}  and CFHTLS-Deep\footnotemark[6] surveys in terms of both area and depth.  Imaging in the $R$ band was done during periods of good seeing (FWHM $\le$ 0.9$^{\prime\prime}$), in an attempt to provide uniform good seeing in one band.  The FWHM of the point-spread-function (PSF) on the $R$ band stacked images is 0.90$^{\prime\prime}$ or better over most of the area, but ranges from 0.76$^{\prime\prime}$ to 1.11$^{\prime\prime}$. In the other bands, the FWHM of the stacked-image PSF ranges from 0.9$^{\prime\prime}$ to 1.2$^{\prime\prime}$.  The $R$ band was used for source detection and shape measurement, and the other bands were used for photometric redshifts. At the time of the shear selection performed by W06, however, only 8.6 deg$^2$ of $R$ band imaging were available, with even less coverage in the other filters.  W06 therefore used an R band source selection ($23<R<25$) over $8.6$~deg$^2$.  They measured source shapes using adaptive second moments \citep{bernjar02}, and convolved the shear field with a \citet{fiscerhtyson97} kernel (inner cutoff 4.25$^\prime$, outer cutoff 50$^\prime$) to obtain convergence maps.
The candidates were selected by their shear ranking and to be withing $5$\amin\ of the survey edge. 
The detection significance of the lowest ranked shear peaks is 3.7$\sigma$ \citepalias[see][ and references within]{wittman06}. 

Our shear-selected sample comes from the \xmm\ follow-up of DLS shear peaks.
We discuss the X-ray observations of eight shear peaks in total. 
Five of them are highly ranked in shear and corresponded to X-ray clusters that had enough signal-to-noise in early \chandra\ follow-up \citepalias{wittman06} to be awarded deep \xmm\ observations to determine X-ray properties.
We add three more DLS shear peaks that go lower into the distribution of shear than went the \citeyear{wittman06} publication; these were awarded shallower \xmm\ observations to confirm as clusters.
The \xmm\ observations (PI:~J.~P.~Hughes) are listed in Table~\ref{tab-obs_info} with their observation identifiers (obsIDs) and exposures. 
Our follow-up naturally divides here into two subsets, as the five shear peaks from the \citeyear{wittman06} paper, hereafter referred to as the original subset, are observed at greater depth in the X-ray  ($\left<t_{exp}\right>=22$ks) than the remaining three ($\left<t_{exp}\right>=10$ks), hereafter called the \emph{beyond} subset. 
For these two subsets we determine the X-ray properties, and for the \emph{beyond} subset, we additionally report the association of the X-ray clusters to the shear peaks.

\begin{figure}
\centering
\includegraphics[width=\columnwidth]{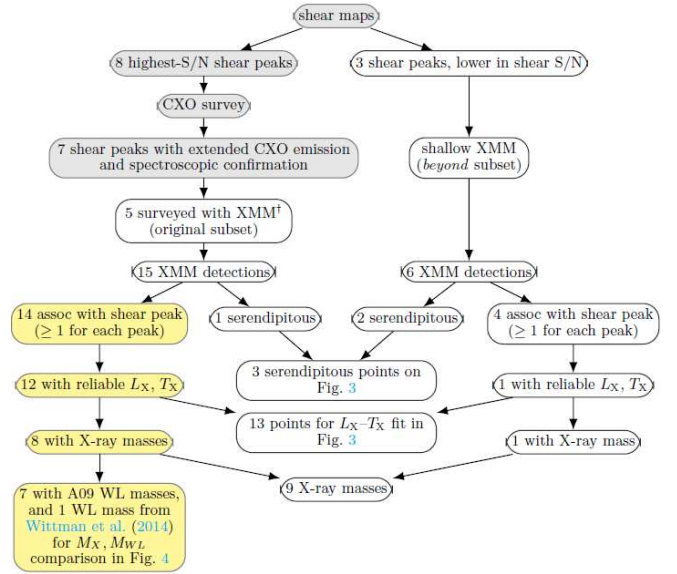}
\figcaption{% 
Subdivision of our shear-selected clusters by X-ray properties. 
Grey shaded boxes at the top differentiate the work completed in \citetalias{wittman06}. 
Yellow shading in the lower left branch highlights the information that results in our mass comparison (\S\ref{subsec-mxm-comp}, Figure~\ref{fig-mm}). 
Subdividing categories for individual clusters are given in Table~\ref{tab-det_info}.
\citetalias{abate09} in the yellow branch refers to \citet{abate09}\\
$^\dagger$ See caption in Table~\ref{tab-obs_info}.%
\label{fig-samp_brkdwn}%
}
\end{figure}

Nearly every shear peak has associated with it more than one X-ray cluster, a likely consequence of the high degree of smoothing in the DLS shear maps.
Some of the clusters that are farther from the shear peaks are detected at lower significance in the X-ray and so we cannot determine the full set of X-ray properties for all of them. 
For clarity, we include a diagram in Figure~\ref{fig-samp_brkdwn} which shows how the X-ray clusters belonging to the original and \emph{beyond} subsets subdivide according to the properties we are able to determine for them. 
Also referenced, in the diagram, are serendipitous X-ray clusters that we find in the observations; these are clusters that could not be confidently associated to the shear peaks. 
We describe, next, our identification and detection of the X-ray clusters, beginning with the imaging required to do so. 

%%%%%%%%%%%%%%%%%%%%%%%%%%%%%%%%%
%	FIGURE : samp_brkdwn
%%%%%%%%%%%%%%%%%%%%%%%%%%%%%%%%%

%%%%%%%%%%%%%%%%%%%%%%%%%%%%%%%%%%%%%%%%%%%%%%%%%%%%%%%%%%%%%%%%%%%%%%%
%	sub section - Imaging
%%%%%%%%%%%%%%%%%%%%%%%%%%%%%%%%%%%%%%%%%%%%%%%%%%%%%%%%%%%%%%%%%%%%%%%

\subsection{Imaging}

We generated images in the soft 0.5-2.0~keV band using \xmm\ data products available through the \emph{XMM-Newton Pipeline Processing System} (\emph{XMM$-$PPS}).  
In particular, we co-added the 0.5-1.0~keV and 1.0-2.0~keV band images, background maps, and exposure maps respectively, and from all three cameras to create a single background-subtracted, exposure-corrected image per observation.   
When relevant, we co-added our 0.5-2.0~keV images from multiple observations (obsIDs) resulting in one soft-band \xray\ image per DLS shear peak.  
\xray\ counterparts were identified on these images, and they were also used to specify regions for spectral extraction. 
These processing steps are described in the following sections.

%%%%%%%%%%%%%%%%%%%%%%%%%%%%%%%%%%%%%%%%%%%%%%%%%%%%%%%%%%%%%%%%%%%%%%%
%	sub section - Source Detection
%%%%%%%%%%%%%%%%%%%%%%%%%%%%%%%%%%%%%%%%%%%%%%%%%%%%%%%%%%%%%%%%%%%%%%%

\subsection{Source Detection} \label{subsec-SourceDetect}

%%%%%%%%%%%%%%%%%%%%%%%%%%%%%%%%%
%	TABLE : det_info
%%%%%%%%%%%%%%%%%%%%%%%%%%%%%%%%%

\begin{table*}
\caption{ X-ray clusters in the \xmm\ observations of DLS shear peaks.}
\resizebox{!}{4cm}{%
\begin{tabular}{ c  r  l  l  r  r | c c c c c}
\hline\\[-3.5ex]
\hline\\[-1.7ex]
No. 		& \multicolumn{1}{c}{Name}	& \multicolumn{1}{c}{X-ray\_\_ID}		& Region		& Rate 		& Signi- & \multicolumn{4}{l}{\it Subdivision by properties} 	\\
 		&   			& 	& $^{\prime}$(kpc)	& ($10^{-3}$cts s$^{-1}$)		&  ficance	& (7) & (8) & (9) & (10) & (11) \\ [2pt]
\hline\\[-2ex]
\multirow{5}{*}[2.5em]{1.}		& \multirow{5}{*}[2.5em]{DLSCL~J0920.1+3029}$^{(a)}$ 	& \phm{(B10a)~}\phm{I}CXOU~J092026+302938	& $3.85$(1034)	& $857 \pm 4$		&	$197$	& $\checkmark$ & $\checkmark$  & $\checkmark$  &  $\checkmark$  &  $\checkmark$ \\
		& 		& \phm{(B10a)~}\phm{I}CXOU~J092053+302800	& $2.57$(672)	& $164 \pm 2$		&	$74$	& $\checkmark$  & $\checkmark$  & $\checkmark$  &  $\checkmark$ &  $\checkmark$ \\
		& 		& \phm{(B10a)~}\phm{I}CXOU J092110+302751	& $2.27$(761)	& $53 \pm 2$		&	$33$	& $\checkmark$  &  $\checkmark$ & $\checkmark$  &  $\checkmark$ &  $\checkmark$ \\
		& 		& \phm{(B10a)~}\phm{I}CXOU~J092011+302954$^{(b)}$	& \nodata		& \nodata		&	\nodata	&  no & no & no  & no & no \\
  		& 		& \phm{(B10a)~}XMMU~J091935+303155 		& $2.17$(728)	& $116 \pm 2$		&	$62$	&  $\checkmark$ &  $\checkmark$ & $\checkmark$  &  $\checkmark$ & $\checkmark$ \\
%-------------------------------------------------------------------------------------------------------------------------------------------------------------
\multirow{4}{*}[1.9em]{2.} 		& \multirow{4}{*}[1.9em]{DLSCL~J0522.2$-$4820} 	& \phm{(B10a)~}\phm{I}CXOU~J052215$-$481816	& $2.19$(580)	& $443 \pm 9$		&	$47$	&  $\checkmark$ &  $\checkmark$ &  $\checkmark$ &  $\checkmark$ &  $\checkmark$ \\ 
		& 		& \phm{(B10a)~}\phm{I}CXOU~J052159$-$481606	& $1.60$(424)	& $87 \pm 6$		&	$15$	&  $\checkmark$ &  $\checkmark$ &  $\checkmark$ & no & no \\ 
		& 		& \phm{(B10a)~}\phm{I}CXOU~J052147$-$482124	& $0.67$(177)	& $3.9 \pm 1.4$		&	$3$	&  $\checkmark$ &  $\checkmark$ & $\checkmark$  & no & no \\
		& 		& \phm{(B10a)~}\phm{I}CXOU~J052246$-$481804	& $1.17$(241)	& $20 \pm 3$		&	$7$	&  no &  $\checkmark$ & no  & no & no \\ 
%-------------------------------------------------------------------------------------------------------------------------------------------------------------
4.		& DLSCL~J1054.1$-$0549 	& \phm{(B10a)~}\phm{I}CXOU~J105414$-$054849 	& $1.25$(238)	& $32 \pm 1$		&	$23$	&  $\checkmark$ &  $\checkmark$ & $\checkmark$  &  $\checkmark$ & $\checkmark$  \\
\\ 
%-------------------------------------------------------------------------------------------------------------------------------------------------------------
\multirow{3}{*}[1.2em]{7.}		& \multirow{3}{*}[1.2em]{DLSCL~J0916.0+2931} 	& \phm{(B10a)~}\phm{I}CXOU~J091551+293637 	& $1.30$(491)	& $17 \pm 1$		&	$12$	&  $\checkmark$ &  $\checkmark$ & $\checkmark$  & no & no \\
  		& 		& \phm{(B10a)~}\phm{I}CXOU~J091601+292750 	& $1.08$(408)	& $37 \pm 2$		&	$22$	&  $\checkmark$ &  $\checkmark$ & $\checkmark$  & $\checkmark$ &  $\checkmark$  \\ 
  		& 		& \phm{(B10a)~}\phm{I}CXOU~J091554+293316$^{(c)}$ 	& \nodata	        & \nodata		        &	\nodata	&  no & no & no  & no & no \\
%-------------------------------------------------------------------------------------------------------------------------------------------------------------
\multirow{2}{*}[0.5em]{8.}		& \multirow{2}{*}[0.5em]{DLSCL~J1055.2$-$0503}	&  \phm{(B10a)~}\phm{I}CXOU~J105535$-$045930	& $1.00$(404)	& $23 \pm 1$		&	$20$	 &  $\checkmark$ &  $\checkmark$ & $\checkmark$  &  $\checkmark$ & $\checkmark$  \\ 
		& 		&  \phm{(B10a)~}\phm{I}CXOU~J105510$-$050414	& $1.26$(534)	& $28 \pm 1$		&	$22$	&  $\checkmark$ &  $\checkmark$ &  $\checkmark$ & no & no \\ [2pt]
%-------------------------------------------------------------------------------------------------------------------------------------------------------------
\multicolumn{6}{l}{\emph{Beyond the initial \citet{wittman06} publication}: } \\[5pt]
\multirow{2}{*}[.5em]{$\;\;\;$B$9.\;\;$} & \multirow{2}{*}[.5em]{DLSCL~J1048.5$-$0411}	& (B9a)\phm{0}~XMMU~J104817$-$041233		& $2.10$(488)	& $77 \pm 3$		&	$22$	&  $\checkmark$ &  $\checkmark$ &  $\checkmark$ &  $\checkmark$  & no\\ 
		& 		& (B9b)\phm{0}~XMMU~J104806$-$041411   	& $0.83$(222) 	& $17 \pm 2$  		&    	$11$	& no &  $\checkmark$ & no  & no & no \\
\multirow{3}{*}[1.2em]{$\;\;\;$B$10.$}	& \multirow{3}{*}[1.2em]{DLSCL~J0921.4+3013}	& (B10a)~XMMU~J092124+301324	& $0.60$(n/a) 	& $5.0 \pm 1.6$		&	$3$ 	&  no &  no &  no & no & no \\
		& 		& (B10b)~XMMU~J092118+301156	& $0.66$(n/a) 	& $4.4 \pm 1.9$		&	$2.4$ 	&  no &  no &  no & no & no \\
		& 		& (B10c)~XMMU~J092102+300530		& $0.88$(332)	& $15 \pm 2$ 		&	$10$ 	& no & $\checkmark$  &  no & no & no\\
$\;\;\;$B$11.$	& DLSCL~J0916.3+3025	& (B11a)~XMMU~J091607+302724		& $1.17$(486) 	& $21 \pm 2$ 		&	$10$ 	&  $\checkmark$ &  $\checkmark$ & no$^{(d)}$  & no & no \\ [2pt]
\hline
\multicolumn{6}{r}{Totals: } & 14 & 17 & 13 & 9 & 8 \\
\hline
\end{tabular}
} % closes resizebox
\tablecomments{ %
The \emph{beyond} subset parenthetical labels are referenced in \S\ref{subsec-SourceDetect}.  
A (n/a) placed where no physical radius can be determined due to lack of redshift.
Column (7) marks the X-ray clusters that \emph{can} be confidently associated to DLS shear peaks (\S\ref{subsec-SourceDetect}). 
Column (8) marks clusters with sufficient statistics to constrain $L_X$ or $T_X$ (see Table~\ref{tab-fit_res} and  \S\ref{subsec-prop}).
Column (9) marks the clusters included in the $L_X-T_X$ fit.
Column (10) marks the clusters for which an X-ray mass could be determined (\S\ref{subsec-xmass}). 
Column (11) marks the clusters with both X-ray and weak lensing masses (\S\ref{subsec-wl-masses}).
\\ $^{(a)}$ This shear peak is coincident with Abell 781, which W06 showed to be a complex of resolved individual clusters.\\
$^{(b)}$ Subcluster of the main cluster of Abell~$781$ \citep{sehg08};  emission is confused with the main component.\\
$^{(c)}$ Central of three X-ray counterparts to DLSCL~J$0916.0$+$2931$; emission is heavily confused with a known point source. \\
$^{(d)}$ Not included in fit because $T_X$ could not be constrained (see Table~\ref{tab-fit_res}).
}
\label{tab-det_info}
\end{table*}%

%%%%%%%%%%%%%%%%%%%%%%%%%%%%%%%%%
%	FIGURE : all_xray
%%%%%%%%%%%%%%%%%%%%%%%%%%%%%%%%%

\begin{figure*}
\centering
\vspace{-2ex}
    \begin{subfigure}[t]{\textwidth}
    \centering \fbox{
        \includegraphics[width=.66\linewidth]{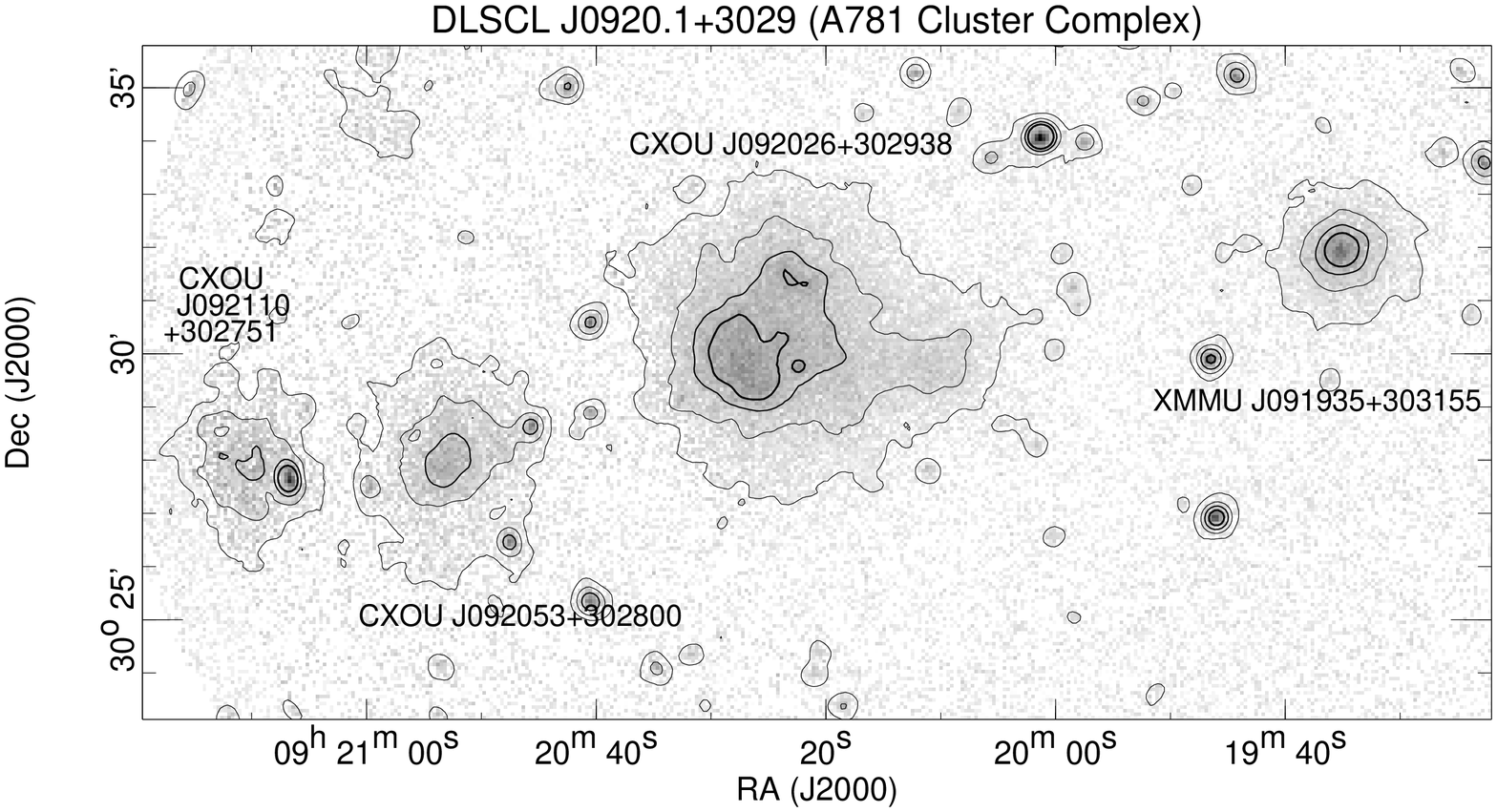} }
    \end{subfigure}
   \centerline{ \fbox{ 
    \begin{subfigure}[t]{0.31\textwidth}
        \centering
        \includegraphics[width=\linewidth]{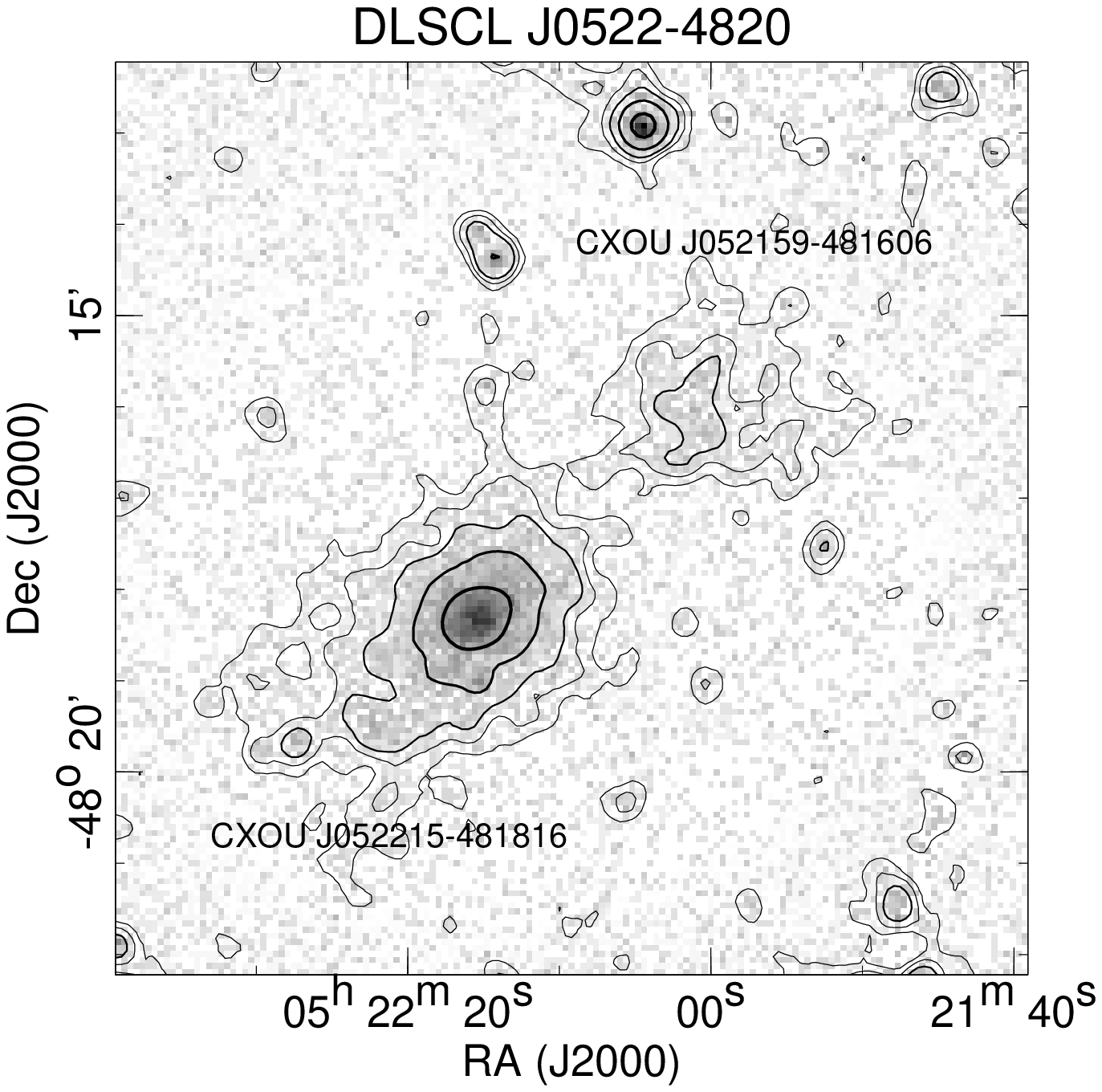} 
    \end{subfigure}%
    \begin{subfigure}[t]{0.31\textwidth}
        \centering
        \includegraphics[width=\linewidth]{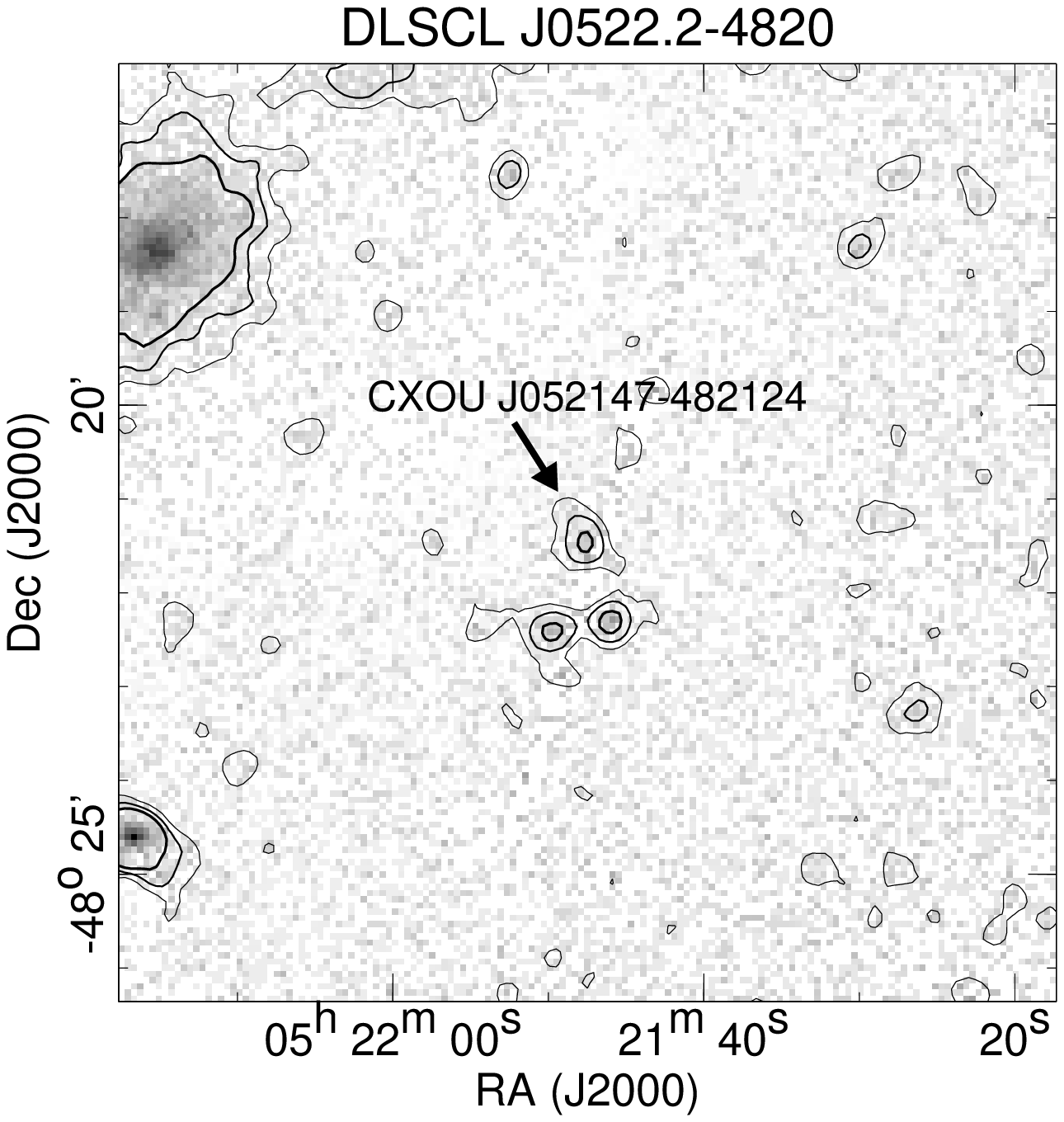} 
    \end{subfigure} } \fbox{% 
    \begin{subfigure}[t]{0.31\textwidth}
        \centering
        \includegraphics[width=\linewidth]{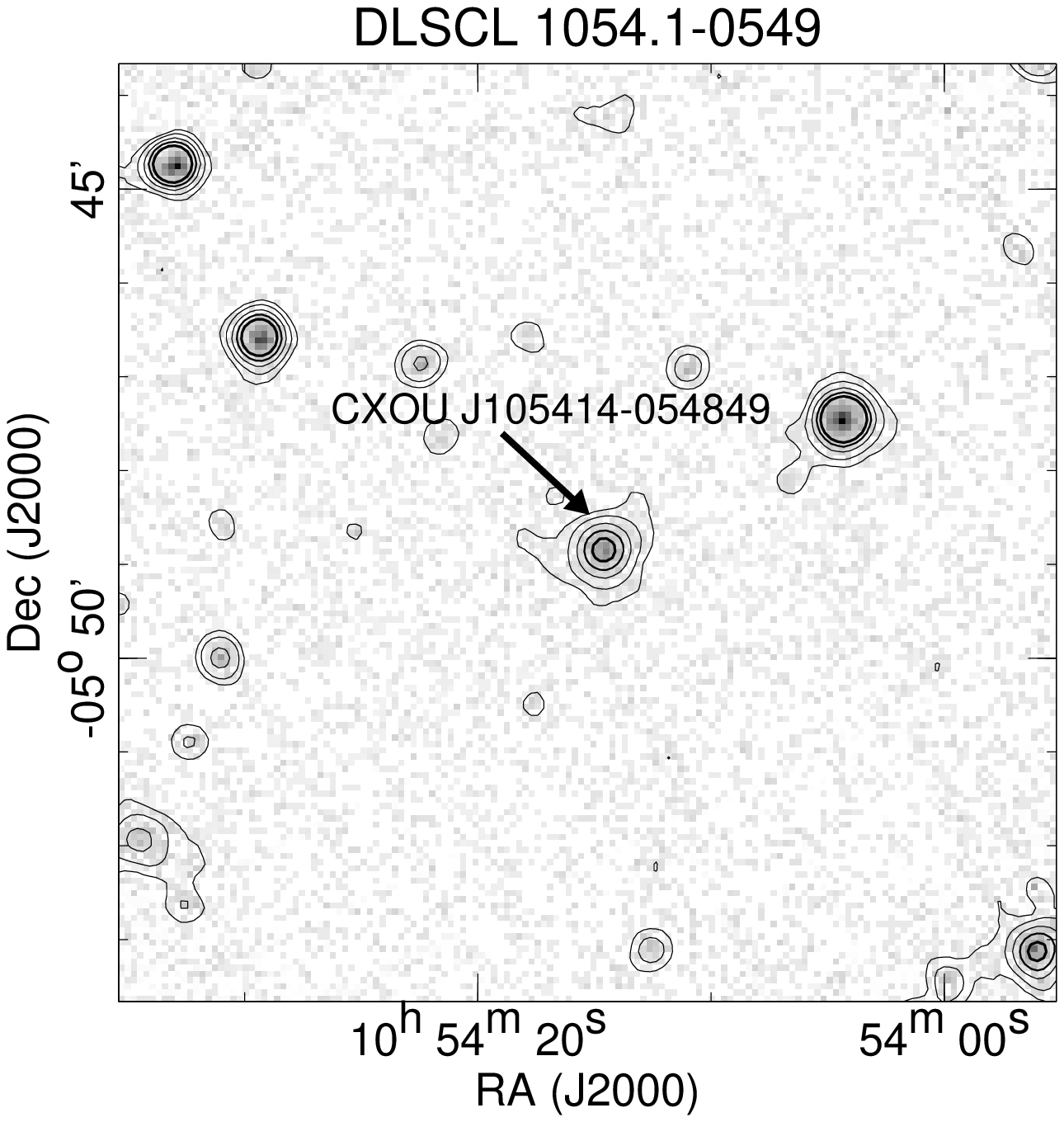} 
    \end{subfigure}    
    } }
   \centerline{\fbox{\hspace{-12ex}
    \begin{subfigure}[t]{0.5\textwidth}
        \centering
        \includegraphics[width=0.66\linewidth]{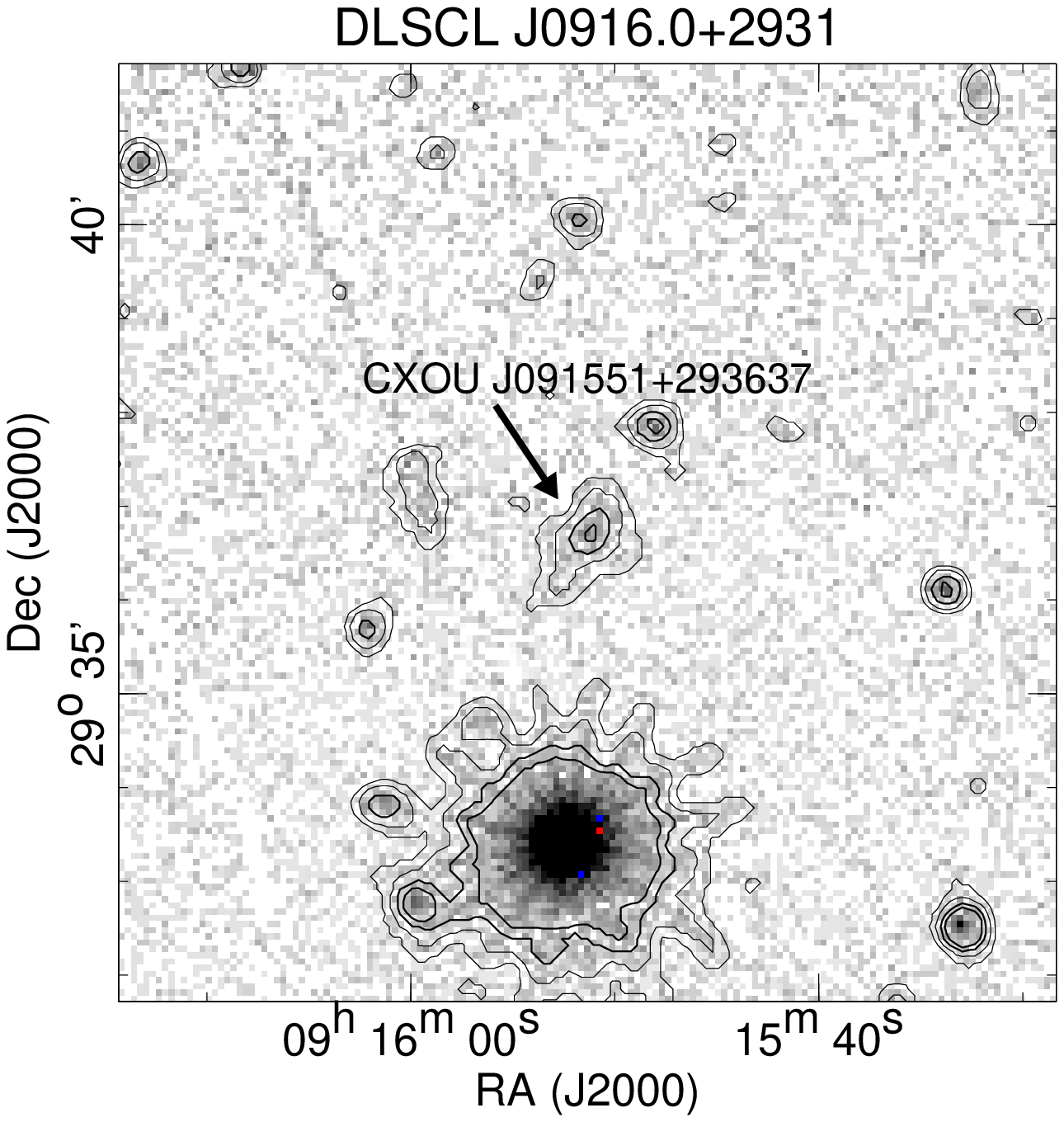} 
    \end{subfigure}%
     \hspace{-25ex}%
    \begin{subfigure}[t]{0.5\textwidth}
        \centering
        \includegraphics[width=0.66\linewidth]{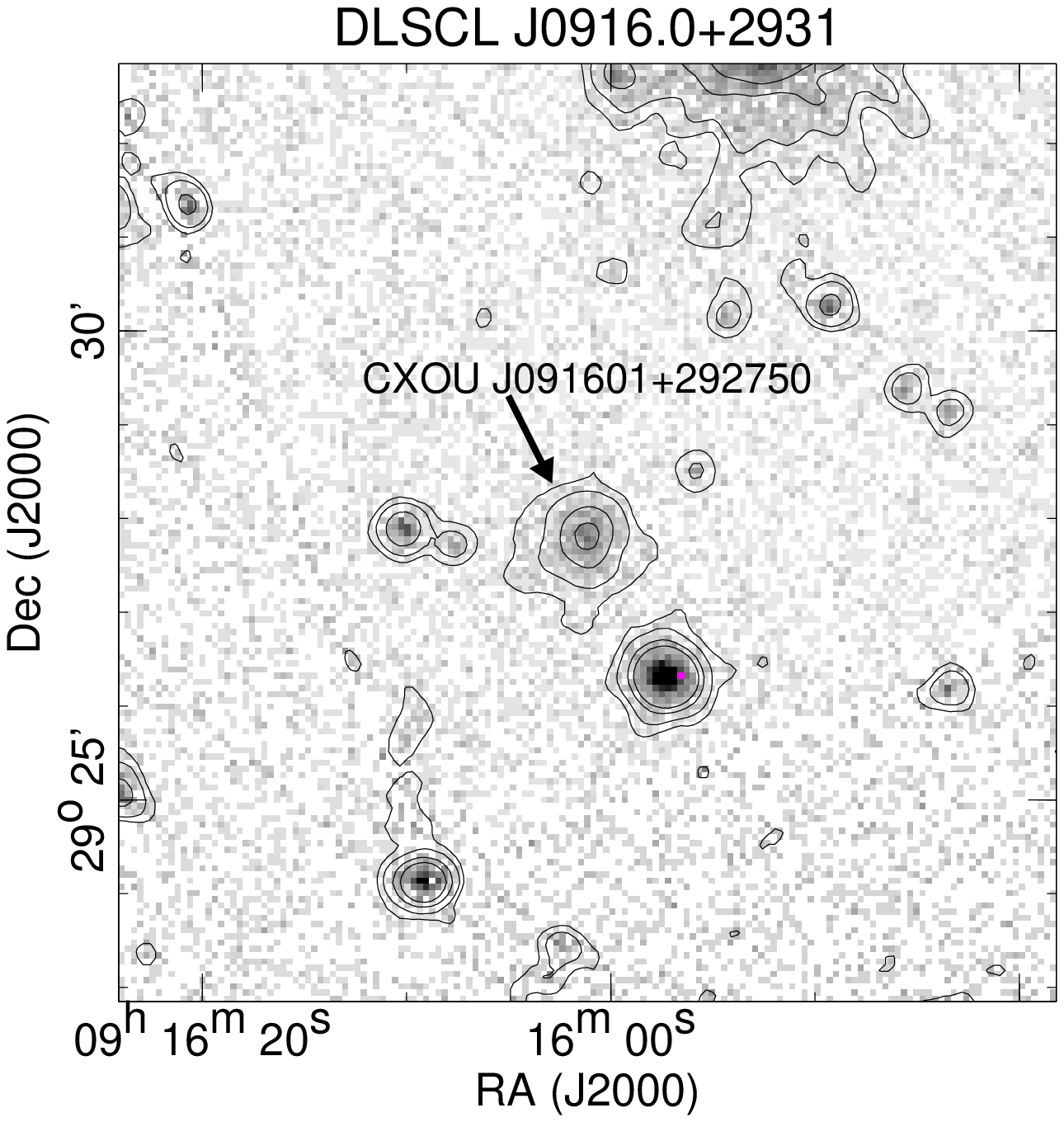} 
    \end{subfigure}
    \hspace{-12ex}}}
\vspace{-2ex}
\figcaption{%
{\it XMM-Newton} images of significant X-ray clusters (labeled within each panel) associated with DLS shear peaks (labels above each panel). 
Panels within a framing box show clusters associated with a single shear peak.
Contour levels were chosen to be at or greater than 1.5 x background level, with an even spacing chosen to show between 3-5 contours per cluster.
The top panel shows the clusters in shear peak \#1, the middle row shows shear peaks \#2 (left and middle) and \#4 (right), and the bottom row shows shear peak \#7. In the bottom
left panel for cluster CXOU~J091551+293637, the prominent object in the bottom of the panel is the point source that confuses emission of the central counterpart to DLSCL~J0916.0+2931. This figure continues on the next page.}
\vspace{-2ex}
\end{figure*}

\begin{figure*}
\ContinuedFloat
\centering
\centerline{ \fbox{ \hspace{-12ex}
\begin{subfigure}{0.5\textwidth}
	\centering
		\includegraphics[width=0.66\linewidth]{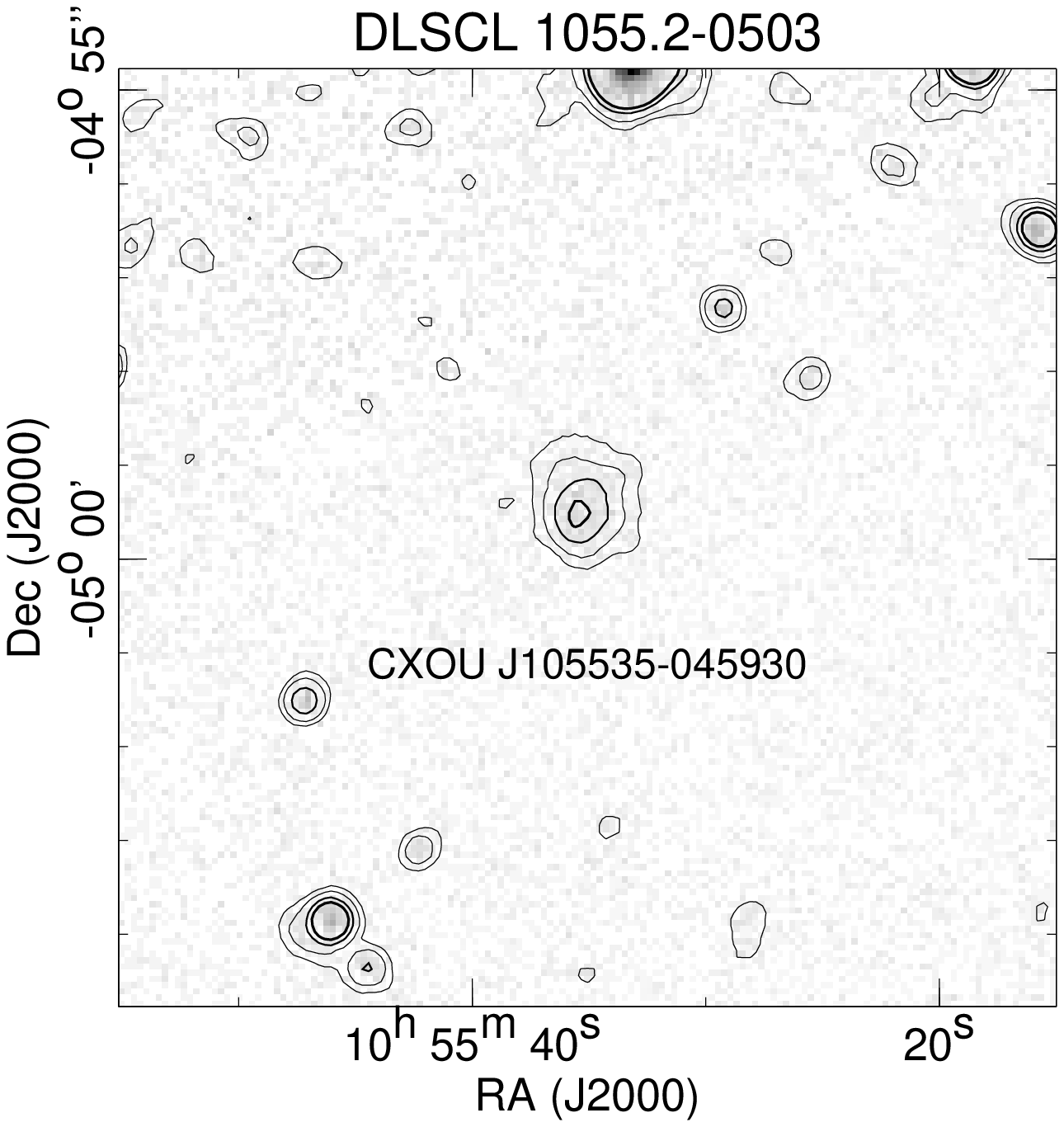}
\end{subfigure} \hspace{-25ex}
\begin{subfigure}{0.5\textwidth}
	\centering
		\includegraphics[width=0.66\linewidth]{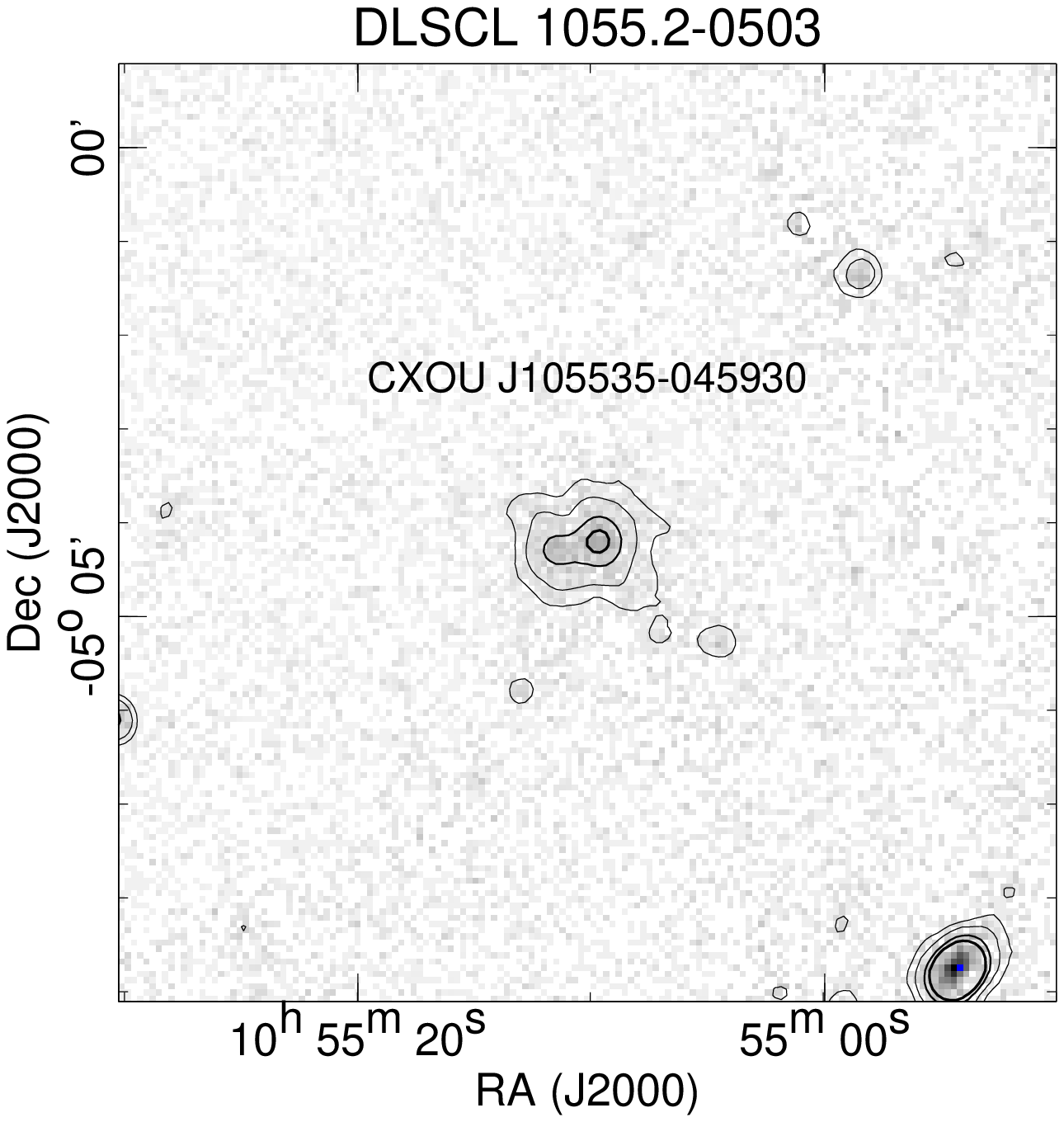}
\end{subfigure}
\hspace{-12ex}} }
\centerline{ \fbox{ \hspace{-12ex}
\begin{subfigure}{0.5\textwidth}
	\centering		\includegraphics[width=0.66\linewidth]{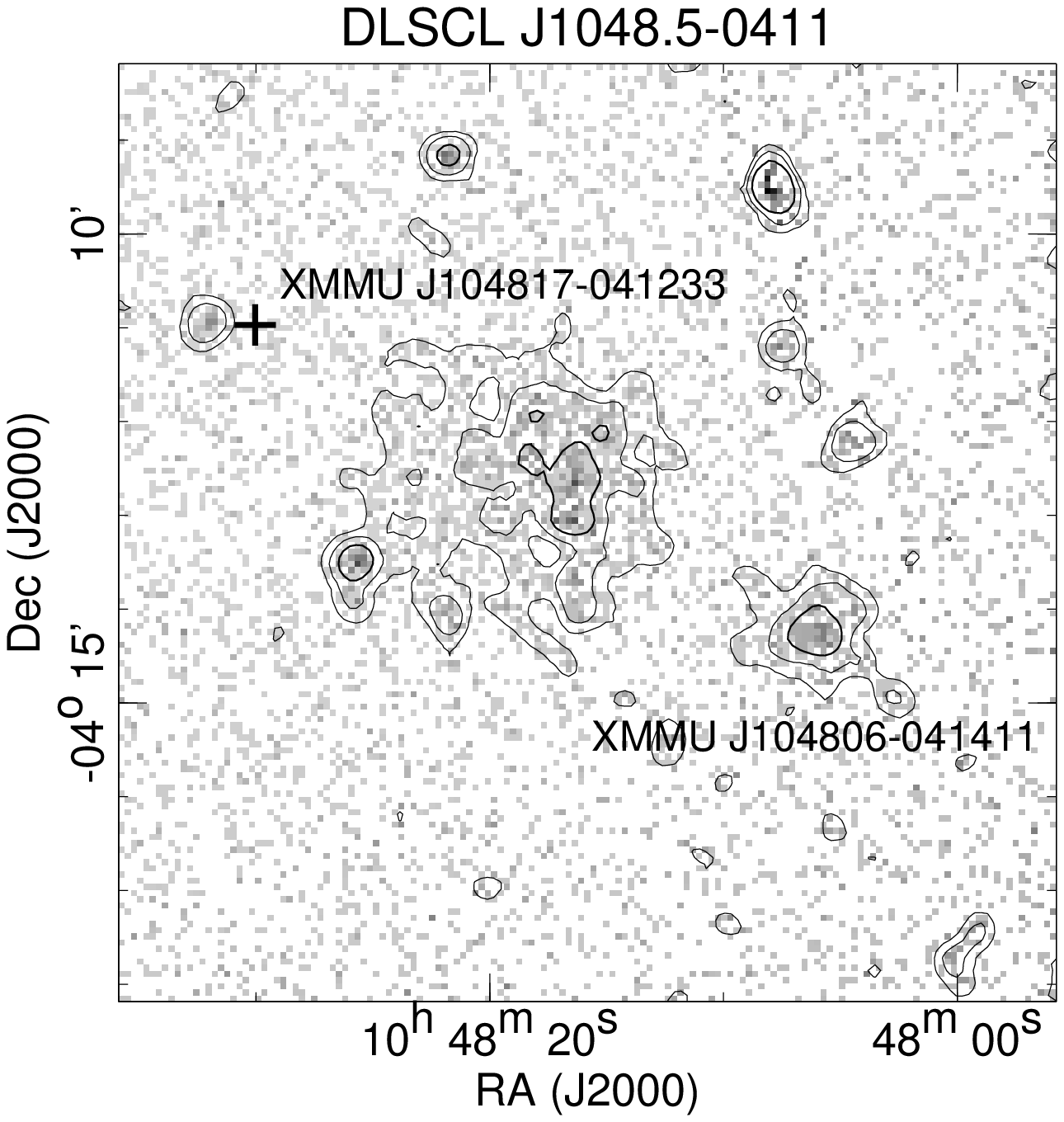}
\end{subfigure} \hspace{-13.5ex}} \hspace{-1ex}  \fbox{ \hspace{-12.5ex}
\begin{subfigure}{0.5\textwidth}
	\centering
		\includegraphics[width=0.66\linewidth]{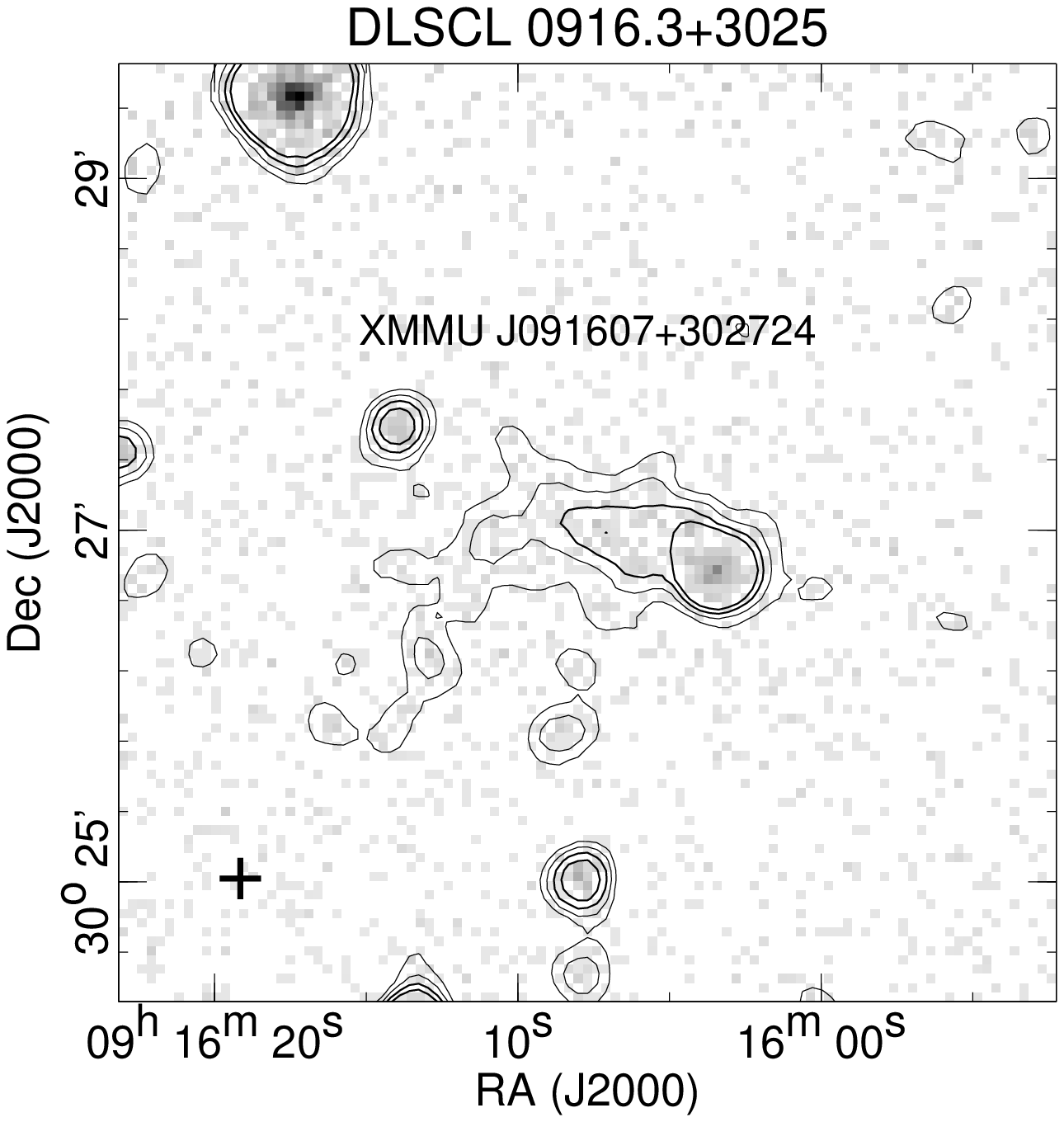}
\end{subfigure}
\hspace{-13.5ex}} }
\figcaption{%
(Continued) 
The top row of panels shows the clusters in shear peak \#8 and the bottom row shows shear peaks \#B9 (left) and \#B11 (right).  The "plus" sign in each of the bottom panels indicates the location of the shear peak.
The cluster in the lower-right portion of the B9 panel cannot be confidently associated with the shear peak, but is labeled here for clarity and cross-referencing with Table~\ref{tab-det_info} and the discussion in section~\ref{subsec-SourceDetect}.
{\it Chandra} imaging (not shown) indicates the presence of a point source near the center of cluster CXOU~J105510-050414 (top right panel). A point source to the west of center is apparent in cluster XMMU~J091607+302724 (bottom right panel).
\label{fig-xall}%
}
\end{figure*}

We recover \xmm\ emission from nearly all of the \xray\ clusters that were identified in \citetalias{wittman06} and were associated to the DLS shear peaks. 
See Table~\ref{tab-det_info} for a list. 
Two of these could not be included in our analysis due to contamination of their X-ray signal. 
The central counterpart to DLSCL~$0916.3$+$2931$, CXOU~J$091554$+$293316$, is heavily confused with a known point source (in the wings of the \xmm\ point-spread-function, see Figure~\ref{fig-xall}).
The emission of the subcluster of Abell~781 \citep[CXOU~J092011+302954, \citetalias{wittman06};][]{sehg08}, is confused with the main cluster's emission with which it is likely merging.
Excluding these two, the detection properties of the remaining recovered sources are given in Table~\ref{tab-det_info}, several of which are imaged in Figure~\ref{fig-xall}.

For the three \emph{beyond} subset shear peaks, we identify potential X-ray counterparts by using the \emph{XMM$-$PPS}.  
On the raw data with updated calibration, we re-run the \xmm\ pipeline which performs its own wavelet decomposition-based source detection.  
The resulting source list is a combined list from source detection performed in multiple bands (soft, and hard) from each camera.  
We verify the extended sources from this list by eye on our soft band images and list them in Table~\ref{tab-det_info} as potential counterparts along with their detection properties.

In the rest of this section we discuss the association of these potential X-ray counterparts to the \emph{beyond} subset shear peaks, referencing any available optical information (from the DLS, or elsewhere). 
The shear peaks in this subset were identified in early work with the DLS (around 2002) and we targeted them for \xmm\ observation; however, they did not make the cut for inclusion in  \citetalias{wittman06}.  
The most significant \xray\ detection in the \emph{beyond} subset is associated with DLSCL~J$1048.5$-$0411$, which is previously unpublished; we discuss its association in the next paragraph. 
The remaining two \emph{beyond} subset shear peaks have appeared previously in the literature:  they are located in DLS field F2, which has been repeatedly studied with new observations in different wave-bands and new weak lensing analyses.
We include these in the context of associating the shear to the X-rays further below.

Near shear peak DLSCL~J$1048.5$-$0411$ of the {\em beyond} subset, there are two extended \xray\ sources detected at high significance, located \about$3.5$\amin\ (B9a) and \about$7$\amin (B9b) away toward the southwest of the DLS position (see the DLSCL J1048.5$-$0411 panel of Fig.~\ref{fig-xall}).
The emission of the former (nearby) \xray\ source lies in an extended high shear region which supports their likely association despite the large offset between the peaks. 
Visual inspection of the DLS data reveals an optical cluster with a brightest cluster galaxy (BCG) that is well centered on the X-ray peak.  
We obtained redshifts of galaxies near this BCG as part of the campaign described in \citetalias{wittman06}.  
We observed the cluster with the Low-Resolution Imaging Spectrograph \citep{oke95} on the Keck~I telescope in April of 2005 and obtained secure redshifts of sixteen galaxies.  
We found eleven galaxies to be likely members, with a mean redshift of $0.2463\pm 0.0006$.
The X-ray emission of this nearby source fits well to a model of thermal cluster emission at this redshift (see Table~\ref{tab-fit_res}). 

For the second \xray\ source,  $\sim7$\amin from the shear peak, the optical association in the DLS is less clear.
A bright, extended, elliptical galaxy rests close to the X-ray position, and is a good candidate for the BCG.  
The \xray s fit well to the emission model of a cluster at the photometric redshift of this galaxy, $z_{ph} = 0.3$.
There are few associated galaxies however, and so without more members, spectroscopy would be required to confidently associate this X-ray cluster with either its neighboring cluster (the nearby cluster above). 
So we do not confidently associate this cluster with the shear peak.

The remaining \emph{beyond} subset shear peaks have been previously reported in the literature as weak lensing detections.
DLSCL~J$0921.4$+$3013$ was reported at a significance of 5.4 in the weak lensing reconstruction of DLS field F2 performed by \citet{kubo09}.
It does not appear, however, in the recent weak lensing analysis of Subaru Hyper Suprime-Cam (HSC) observations of F2, conducted by \citet{miyazaki15}. 
The \emph{XMM$-$PPS} finds three extended \xray\ sources in this vicinity: two toward the north (B10a and B10b, Table~\ref{tab-det_info}) and one toward the south (B10c, Table~\ref{tab-det_info}). 

The southern \xray\ source, XMMU~J$092102$+$300530$, is approximately  \about$8$\amin\ to the southeast of the DLS position and has no other weak lensing peak nearby. 
Thus, we cannot associate this X-ray source to a DLS shear peak.
There is no corresponding cluster in the optical cluster catalog from the DLS \citep{ascaso14} due to their bright star mask; however, visual inspection of the DLS images shows clear evidence for an optical cluster beyond the offending star. 
We estimate a photometric redshift ($z=0.53$) from the  galaxies in the cluster outskirts. 
The X-rays also fit nicely to a thermal cluster emission model at this redshift (see entry in Table~\ref{tab-fit_res}). 

The two northern \xray\ detections, B10a and B10b, lie closer to the DLS position (\about3.5\amin away).
They are small, $\lesssim1$\amin\ sized clumps, which overlie a much broader region of red galaxies in the DLS at similar redshifts ($z\sim 0.6$). 
\citet{ascaso14} report an optical cluster between the \xray\ clumps at a redshift between $0.54<z<0.6$.
The \xray\ clumps do not have well defined peaks or shapes, and are difficult to associate with the galaxies as independent clusters or as a single cluster with poor \xray\ emission.
We find these data to be consistent with the interpretation presented in \citet{starik14} as a superposition of low mass systems which they suggest after comparing their own X-ray analysis to groups identified in the SHELS \citep[Smithsonian HEctoscopic Lensing Survey:][]{geller10}.  
Furthermore, the detection significance  for the \xray\ sources (XMMU~J092124+301324 and XMMU~J092118+301156) measured within regions sized by eye to maximize the extracted flux (for numerical values see Table~\ref{tab-det_info}), is low and so we do no further spectral analyses on them. 
Because these two are too faint, and the southern cluster (B10c) is too far away, we cannot report properties of associated X-ray clusters for DLSCL~J0921.4+3013.

%%%%%%%%%%%%%%%%%%%%%%%%%%%%%%%%%
%	TABLE : fit_res_det
%%%%%%%%%%%%%%%%%%%%%%%%%%%%%%%%%

\begin{table*}
\caption{\xmm\ spectral fitting results: temperature and luminosity}
\resizebox{0.95\textwidth}{!}{\begin{minipage}{\textwidth}
\begin{tabular}{ c  r  l  r@{/}l  c  c  c  c  c }
\hline\\[-3.5ex]
\hline\\[-1.7ex]
			& \multicolumn{1}{c}{Name}  	& \multicolumn{1}{c}{X-ray\_\_ID$^\ddag$}  	&\multicolumn{2}{c}{$\chi^2$/ $d.o.f.$}	& $nH$  						& $z$ 						& Abund.  					& $kT_X$					& $L_{X}$  						\\
			&    						&  								&\multicolumn{2}{c}{}					& (LAB) 						&  							&  							&						&  $Bolometric$ 					\\ 
 			& \multicolumn{1}{c}{} 		&  								&\multicolumn{2}{c}{}					& $10^{20}$cm$^{-2}$ 		&  							&  $Z_\odot$					& keV					& $10^{44}$ ergs s$^{-1}$  		\\[2pt]
\hline\\[-2ex]
\multirow{4}{*}[1.9em]{1.} 			& \multirow{4}{*}[1.9em]{DLSCL~J0920.1+3029} 	 	& \phm{I}CXOU~J092026+302938		&	$2580$			&	$1881$			&  	$1.65$ 					& $0.302^{\hphantom{(a)}}$					& $0.21^{+0.02}_{-0.02}$ 	& $6.33^{+0.13}_{-0.13}$	& $10.55^{+0.07}_{-0.07}$		\\
  			& 							& \phm{I}CXOU~J092053+302800		&	$1071$			&	$964$			&	$1.65$					& $0.291^{\hphantom{(a)}}$					& $0.21^{+0.05}_{-0.05}$		& $3.19^{+0.13}_{-0.13}$	& $~2.08^{+0.06}_{-0.06}$		\\
  			& 							& \phm{I}CXOU~J092110+302751		&	$659$			&	$494$			&	$1.65$					& $0.427^{\hphantom{(a)}}$					& $0.3$						& $3.87^{+0.41}_{-0.33}$	& $~2.67^{+0.07}_{-0.07}$		\\
   			& 							& XMMU~J091935+303155 			&	$833$			&	$768$			& 	$1.66$					& $0.428^{(a)}$					& $0.3$						& $3.41^{+0.15}_{-0.15}$	& $~3.30^{+0.05}_{-0.05}$		\\
%-------------------------------------------------------------------------------------------------------------------------------------------------------------------------------------------------------------------------------------------------------------------------------
\multirow{4}{*}[1.8em]{2.} 			& \multirow{4}{*}[1.8em]{DLSCL~J0522.2$-$4820}		& \phm{I}CXOU~J052215$-$481816	&	$249$			&	$286$			& 	$2.85$					& $0.296^{\hphantom{(a)}}$					& $0.3$						& $4.03^{+0.25}_{-0.24}$	& $~3.67^{+0.06}_{-0.06}$		\\ 
      		& 							& \phm{I}CXOU~J052159$-$481606	&	$19$				&	$22$				& 	$2.82$					& $0.296^{\hphantom{(a)}}$					& $0.3$						& $4.34^{+1.31}_{-0.87}$	& $~0.84^{+0.05}_{-0.05}$		\\ 
      		& 							& \phm{I}CXOU~J052147$-$482124	&	$3$		 		&	$5$				& 	$2.79$					& $0.296^{\hphantom{(a)}}$					& $0.3$						& $1.05^{+0.44}_{-0.31}$	& $~0.05^{+0.02}_{-0.01}$ 		\\ 
     		& 							& \phm{I}{\it CXOU~J052246$-$481804}$^\Delta$	&	$11$				&	$11$				& 	$2.91$					& $0.210^{\hphantom{(a)}}$					& $0.3$						& $1.48^{+0.44}_{-0.21}$	& $~0.07^{+0.01}_{-0.01}$ 		\\
%-------------------------------------------------------------------------------------------------------------------------------------------------------------------------------------------------------------------------------------------------------------------------------
4.	 		& DLSCL~J1054.1$-$0549		& \phm{I}CXOU~J105414$-$054849 	&	$79$				&	$63$				& 	$2.43$					& $0.190^{\hphantom{(a)}}$					& $0.3$						& $1.07^{+0.03}_{-0.04}$	& $~~0.06^{+0.003}_{-0.003}$		\\
%-------------------------------------------------------------------------------------------------------------------------------------------------------------------------------------------------------------------------------------------------------------------------------
\multirow{2}{*}[0.5em]{7.}      	& \multirow{2}{*}[0.5em]{DLSCL~J0916.0+2931} 		& \phm{I}CXOU~J091551+293637 	&	$17$				&	$15$				& 	$1.72$					& $0.530^{\hphantom{(a)}}$					& $0.3$						& $1.44^{+0.22}_{-0.16}$	& $~0.58^{+0.10}_{-0.10}$		\\
 			& 							& \phm{I}CXOU~J091601+292750		&	$62$		 		&	$55$				&	$1.74$					& $0.531^{\hphantom{(a)}}$					& $0.3$						& $2.09^{+0.19}_{-0.19}$ & $~1.01^{+0.07}_{-0.07}$		\\
%--------------------------------------------------------------------------------------------------------------	-----------------------------------------------------------------------------------------------------------------------------------------------------------------
\multirow{2}{*}[0.5em]{8.} 			& \multirow{2}{*}[0.5em]{DLSCL~J1055.2$-$0503}		& \phm{I}CXOU~J105535$-$045930	&	$36$				&	$38$				& 	$2.40$					& $0.609^{\hphantom{(a)}}$					& $0.3$						& $3.38^{+0.46}_{-0.44}$	& $~1.04^{+0.07}_{-0.06}$		\\
 			& 							& \phm{I}CXOU~J105510$-$050414	&	$33$				&	$32$				&  	$2.39$					& $0.680^{\hphantom{(a)}}$					& $0.3$ 						& $4.14^{+0.69}_{-0.57}$	& $~2.80^{+0.17}_{-0.17}$		\\[2pt]
%-------------------------------------------------------------------------------------------------------------------------------------------------------------------------------------------------------------------------------------------------------------------------------
\multicolumn{8}{l}{\emph{Beyond the initial \citet{wittman06} publication}: }	\\[2pt]
\multirow{2}{*}[0.55em]{$\;\;\;$B$9.\;\;$} & \multirow{2}{*}[0.55em]{DLSCL~J1048.5$-$0411}		& XMMU~J104817$-$041233			&	$38$				&	$40$	 			& 	$3.69$					& $0.246^{(b)}$ 					& $0.3$						& $2.38^{+0.36}_{-0.29}$	& $~0.55^{+0.03}_{-0.03}$		\\
			& 					& {\it XMMU~J104806$-$041411}$^\Delta$ 			&	$8$				&	$9$				&	$3.69$					& $0.30^{(c)}\hphantom{0}$					& $0.3$						& $1.64^{+0.44}_{-0.27}$	& $~0.17^{+0.02}_{-0.02}$		\\
$\;\;\;$B$10.$	& DLSCL~J0921.4+3013		& {\it XMMU~J092102+300530}$^\Delta$			&	$13$				&	$13$				&	$1.65$					& $0.53^{(c)}\hphantom{0}$					& $0.3$						& $2.08^{+0.65}_{-0.43}$	& $~0.56^{+0.06}_{-0.06}$		\\
$\;\;\;$B$11.$	& DLSCL~J0916.3+3025			& {\it XMMU~J091607+302724}			&	$16$				&	$20$				&	$1.10$					& $0.650^{(d)}$					& $0.3$						& $3^\dagger$ 				& $~1.12^{+0.10}_{-0.10}$	\\[3pt]
\hline
\end{tabular}
\end{minipage}
}
\tablecomments{Redshift sources:}
\tablenotetext{}{$^{(a)}$ \citet{sehg08}, $^{(b)}$ this work - spectroscopy (\S\ref{subsec-SourceDetect}), $^{(c)}$ this work - DLS photometry (\S\ref{subsec-SourceDetect}), and $^{(d)}$ \citet{geller14}.%
}
\tablenotetext{\scriptstyle\ddag}{Italics denote clusters that are not used for the $L_X$-$T_X$ relation for the reasons given by the following footnotes.}
\tablenotetext{\scriptstyle\Delta}{X-ray cluster not confidently associated to shear peak.}
\tablenotetext{\scriptstyle\dagger}{Temperature fixed at nominal value; data could not constrain.}
\label{tab-fit_res}
\end{table*}

The third shear peak in the \emph{beyond} subset, DLSCL~J$0916.3$+$3025$, was not found in the weak lensing analysis of \citet{kubo09}.
More recently, two weak lensing detections near this position have been reported.  
\citet{utsumi14}, in their analysis of a Subaru Suprime camera observation of a part of the DLS field F2, and \citet{miyazaki15}, in their weak lensing analysis of a Subaru Hyper Suprime-Cam (HSC) observation covering all of the same DLS field, both find weak lensing detections that are within \about1\amin\ (but on opposite sides) of the corresponding \xray\ source  position (of B11).
A nearby optical cluster \citep[$z\sim 0.54,$][]{ascaso14} is found to be approximately \about$1.5$\amin\ away from the \xray\ source and possibly consistent with the \citet{miyazaki15} weak lensing peak.
We make a plausible association between the X-ray source and the Miyazaki/Utsumi detections.  
The X-rays are faint and do fit to a thermal cluster emission model, but with the temperature fixed at a nominal value (Table~\ref{tab-fit_res}). 
We describe next our steps to extract spectra and fit them to measure luminosities and temperatures.
 
%%%%%%%%%%%%%%%%%%%%%%%%%%%%%%%%%%%%%%%%%%%%%%%%%%%%%%%%%%%%%%%%%%%%%%%
%	sub section - Extracting Spectra
%%%%%%%%%%%%%%%%%%%%%%%%%%%%%%%%%%%%%%%%%%%%%%%%%%%%%%%%%%%%%%%%%%%%%%%

\subsection{Extracting Spectra} \label{subsec-ExtractingSpectra}

We generated \xray\ spectra from newly calibrated event-lists; these are among the outputs of the \emph{XMM-PPS} run performed above (\S\ref{subsec-SourceDetect}), on the raw data with updated calibration files.
We use the \emph{XMM-Newton Science Analysis System }(\emph{XMM-SAS}) software package (version 11.0) to run this pipeline and to process this data further. 
The newly calibrated event-lists from each camera were filtered in time to remove periods of highly flaring soft proton background.  
The resulting exposures are listed in Table~\ref{tab-obs_info}.  
Spectra and other products necessary for spectral fitting were generated with these flare-filtered data.  

Spectra were extracted from within regions that we defined on our soft-band images. 
These same regions were also used to determine the detection properties and are listed alongside in Table~\ref{tab-det_info}.
We began the region selection by drawing contours on our $0.5-2.0$~keV images, at levels of count rate per pixel that are 1.5 times the background level and higher.  
The outermost contour guided our initial choice of either circular or elliptical source region, which was placed to just surround the contour.  
We adjusted the size of this region by 5 or 10 percent iteratively until the luminosity measured from within converged.  
In this way, we were sure to have collected all of the cluster emission with minimal contamination from the unresolved X-ray background.  
Resolved background sources, found either by \emph{XMM-PPS} or present obviously in available \emph{Chandra} images, were excluded.  
Background regions were placed as annuli around source regions, and also excluded any \emph{XMM-PPS} detected point sources or neighboring cluster regions.  

Spectra and other data products used in fitting (arfs and rmfs) were generated with the standard binnings, event filters, and other recommended parameters suggested by the \xmm\ team for analyzing extended X-ray sources. 
Among these recommendations, we chose to weight the response files by the cluster images to better account for brightness variations.  

%%%%%%%%%%%%%%%%%%%%%%%%%%%%%%%%%%%%%%%%%%%%%%%%%%%%%%%%%%%%%%%%%%%%%%%
%		sub section - Temperature and Luminosity
%%%%%%%%%%%%%%%%%%%%%%%%%%%%%%%%%%%%%%%%%%%%%%%%%%%%%%%%%%%%%%%%%%%%%%%

\subsection{X-ray Temperature \& Luminosity} \label{subsec-prop}

\xray\ spectra from regions described above (\S\ref{subsec-ExtractingSpectra}) were fit in \xspec\ to a product of the \mekal\ model \citep{mew85,mew86,kaas92} and the {\tt phabs} model.  
The \mekal\ model describes a thermal plasma with ionized atomic components, with model parameters describing gas temperature, abundance, redshift, and the emission measure (proportional to the fit normalization).  
The {\tt phabs} model describes galactic photoelectric absorption and depends only on one parameter, the absorbing column density.

We generally let temperature and normalization vary, and fixed all remaining parameters. 
We fixed abundance to 0.3~$\left[\textup{Z}_\odot\right]$ except when data quality could support a constraint. 
The redshift was set to the spectroscopic value determined by the DLS \citepalias{wittman06} or other follow up work (indicated in Table~\ref{tab-fit_res}).  
If the data were too poor to constrain both temperature and normalization, we fixed the temperature to a reasonable value (see Table~\ref{tab-fit_res}).  
In all cases, the column density for the {\tt phabs} model was fixed to the galactic neutral hydrogen column densities measured by the Leiden/Argentine/Bonn (LAB) survey \citep{Kalb05} at the cluster position.  

All spectra of each cluster, one from each of the three cameras, were simultaneously fit to one function in \xspec. 
Uncertainties due to poor subtraction of telescopic fluorescence lines were addressed independently for PN and MOS by excluding the affected channels.  
Background scaling was adjusted by examining the high energy [10~keV$-$12~keV] counts \citep{vikh09} where no source emission is expected. 
High energy channels, where emission from a given cluster was negligible, were excluded from the spectral fit. 

The resulting temperatures and bolometric luminosities from these fits, for $14$ X-ray clusters associated with DLS shear peaks, and three serendipitous X-ray clusters, are shown in Table~\ref{tab-fit_res}, along with the corresponding model parameters.  
Temperatures and luminosities are also plotted in Figure~\ref{fig-lt}, where the luminosities corrected for expansion (i.e., $\times E(z)^{-1}$).
Previously determined properties for clusters that overlap with other studies \citep[e.g.,][]{sehg08,starik14} are consistent with the values we determine within errors. 
The ranges of these properties are broad, spanning over four orders of magnitude in luminosity and a factor of six in temperature.  

The luminosity range includes the order of the brightest known clusters ($10^{45}$ erg s$^{-1}$), as well as that of small groups ($10^{41}-10^{42}$ erg s$^{-1}$).
The temperature range does not reach very high, but includes the average hot cluster ($\gtrsim 5$~keV) as well as many low group-like values ($\sim 1$keV).
Morphology is difficult to quantify for the whole sample due to some clusters with poor statistics, but a visual examination of the sample (see Fig.~\ref{fig-xall}) reveals a full range from smooth and highly centrally peaked to generally disturbed and lumpy.
The disturbed morphology seems to be associated with both interacting systems (see \S\ref{subsec-xmass}), and isolated ones.
To understand these sample properties in the context of other well-understood \xray\ clusters we make a comparison of the luminosities and temperatures as well as of the $L_X-T_X$ relation to \xray\ selected samples from the literature.

We choose literary comparison samples that are selected in the \xray, and that have luminosities and temperatures determined without excising cores, as we do. 
Two such samples, with comparable redshift, temperature and luminosity ranges, are presented by \citet{maugh12}, and \citet{hilton12}, hereafter called \citetalias{maugh12} and \citetalias{hilton12}.  
We find that the general distribution of cluster morphology is also consistent with those of the \citetalias{maugh12} and \citetalias{hilton12} samples.
We discuss our $L_X-T_X$ relation, and its comparison to the relations derived in \citetalias{hilton12} and \citetalias{maugh12} samples next. 

%%%%%%%%%%%%%%%%%%%%%%%%%%%%%%%%%%%%%%%%%%%%%%%%%%%%%%%%%%%%%%%%%%%%%%%
%	     sub section - Lx-Tx Relation
%%%%%%%%%%%%%%%%%%%%%%%%%%%%%%%%%%%%%%%%%%%%%%%%%%%%%%%%%%%%%%%%%%%%%%%

\subsection{$L_X-T_X$ relation}\label{subsec-lxtx}
The luminosity-temperature relation of typical X-ray clusters is a tight correlation, born out of the cluster growth process \citep[e.g.,][]{kaiser86, vikh09}. 
Historically, it has distinguished samples that deviate from self-similarity \citep[e.g.,][]{mark98,arnaud99}.
More recent studies have shown the slope of this relation to vary with measures of dynamical activity in clusters \citep[\citetalias{maugh12};][]{mahdavi13}.
Clusters selected without regard to dynamical state tend to show higher slopes ($>3$) \citep[e.g.,][]{mantz10,hilton12}, as also do smaller
galaxy groups \citep[e.g.,][]{sun09}.
On the other hand, carefully selected samples of relaxed clusters, with core-removed temperatures \citepalias[e.g.,][]{maugh12}, produce $L_X-T_X$ relations closer to the expected self-similar relation ($L\sim T^2$).
Selection effects also affect the slope, such as e.g., Malmquist bias, which tends to lower the slope. 
We determine the luminosity-temperature relation of our shear-selected sample to see how the clusters scatter around the best fit relation, and to compare its best fit relation to X-ray selected cluster samples.

\begin{figure}
\centering
\includegraphics[width=\columnwidth]{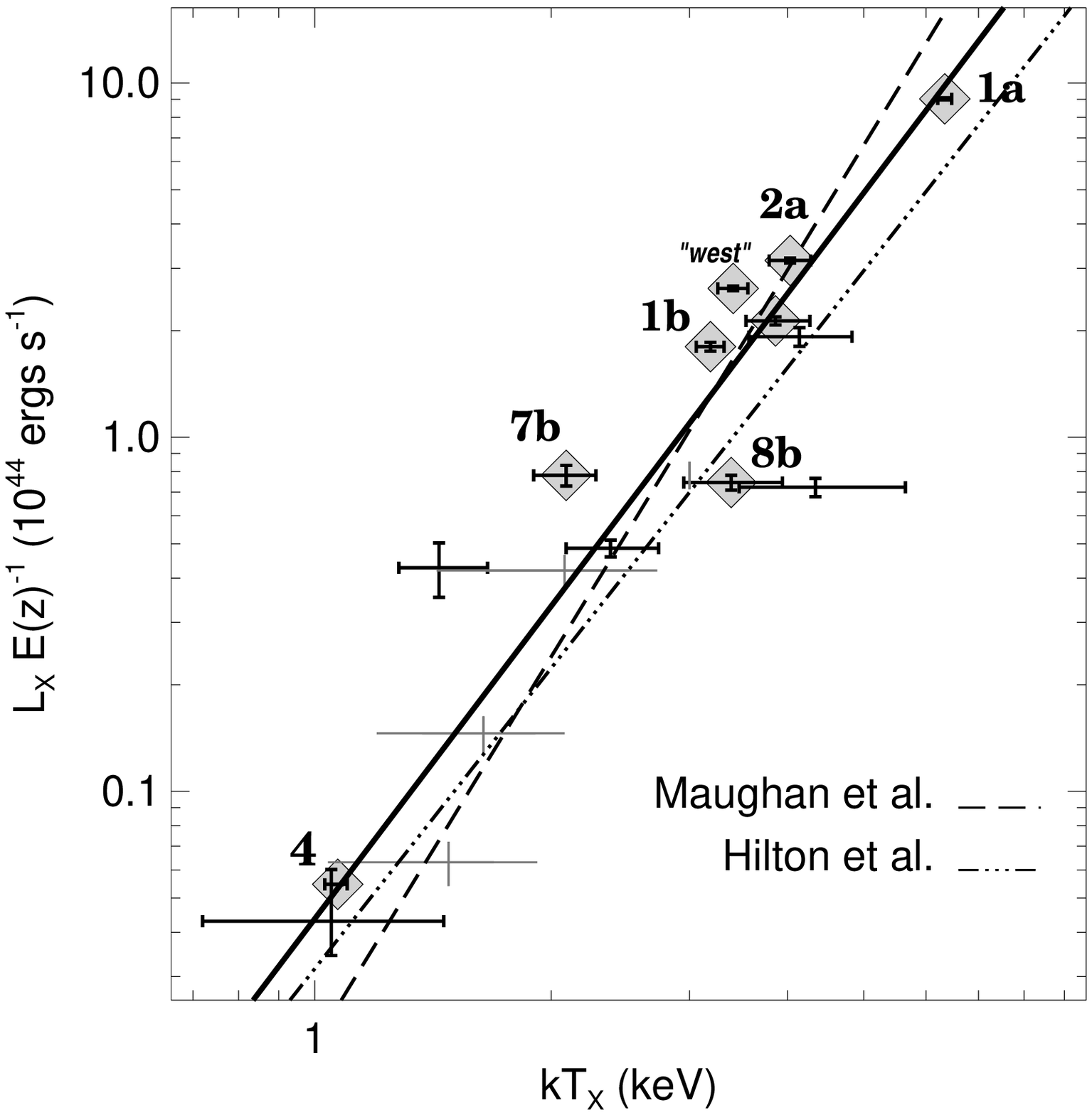}
\figcaption{Temperatures and bolometric luminosities from Table~\ref{tab-fit_res}, plotted as black crosses showing 1 $\sigma$ error bars.  The solid line is the fit to our data.  Filled diamonds mark clusters with both X-ray and weak lensing masses and are given labels from Table~\ref{tab-rmt500}. The un-labeled diamond point is cluster 1c.  Grey points were not included in the fit; these are clusters that we were unable to confidently associate with DLS shear peaks (\S\ref{subsec-SourceDetect})\label{fig-lt}}
\end{figure}

The mathematical form of the relation to which we fit our $L_X$, $T_X$ data is,
\begin{equation}
h^{-1}E(z)^{-1}L_{X,bol} = L_0\left(T_X/5.0 keV\right)^\alpha.
\end{equation}
The data were fit by performing a linear regression on the logarithm of the luminosities and temperatures.  
We use the orthogonal \bces\ method \citep{akritas96} for the regression, after symmetrizing our errors in temperature by averaging the absolute value of the  errors in each direction.  
We choose the BCES method because so do the studies to which we compare our resulting luminosity-temperature relation. 
Our best fit is shown in Figure~\ref{fig-lt} by the solid line; it has a log space slope of $\alpha=2.93 \pm 0.15$ and an intercept of $\textup{log}(L_0 / 1\ \textup{erg s}^{-1}  ) = 44.69 \pm 0.08$.  
X-ray clusters that could not be associated with shear peaks (noted in Table~\ref{tab-fit_res}) were not included in the fit, but are shown on the plot as unmarked grey error bars.  

First we note that our $L_X-T_X$ relation is not consistent with the self-similar slope ($\alpha=2$).
The $L_X$, $T_X$ points scatter tightly around the comparison relations as well as the best fit plotted in Figure~\ref{fig-lt}.
The \citetalias{maugh12} relation (dashed line) is from their full sample of $114$ clusters selected without regard to dynamical state. 
Their slope of $\alpha=3.63$ is slightly steeper than ours, although still consistent at $<$2$\sigma$.
The \citetalias{hilton12} relation (dash-triple dotted line) is from their intermediate redshift ($0.25 < z < 0.5$) sample of $77$ clusters also selected without regard to dynamical state; it has a slightly shallower slope than ours at $\alpha=2.82$.  
Both of these relations for X-ray selected samples are consistent with our clusters selected in weak lensing.  
This result is also consistent with and confirms a similar finding by \citet{giles15}, who fit nine shear-selected X-ray clusters to an $L_X-T_X$ relation of similar form and get a slope and normalization $\alpha = 2.63\pm 0.69$ and $\log( L_0 / 1 \textup{ erg s}^{-1} ) = 44.44 \pm 0.15$ (we have converted their normalization to one that would match a pivot-temperature of $5$~keV). 

%%%%%%%%%%%%%%%%%%%%%%%%%%%%%%%%%%%
%        Figure $L_X-T_X$
%%%%%%%%%%%%%%%%%%%%%%%%%%%%%%%%%%%

%%%%%%%%%%%%%%%%%%%%%%%%%%%%%%%%%%%%%%%%%%%%%%%%%%%%%%%%%%%%%%%%%%%%%%%
%%%%%%%%%%%%%%%%%%%%%%%%%%%%%%%%%%%%%%%%%%%%%%%%%%%%%%%%%%%%%%%%%%%%%%%
% %
% %		SECTION - Masses
% %
%%%%%%%%%%%%%%%%%%%%%%%%%%%%%%%%%%%%%%%%%%%%%%%%%%%%%%%%%%%%%%%%%%%%%%%
%%%%%%%%%%%%%%%%%%%%%%%%%%%%%%%%%%%%%%%%%%%%%%%%%%%%%%%%%%%%%%%%%%%%%%%

\section{Mass Estimates \& Comparison} \label{sec-mass}

%%%%%%%%%%%%%%%%%%%%%%%%%%%%%%%%%
%	TABLE : rmt500
%%%%%%%%%%%%%%%%%%%%%%%%%%%%%%%%%
\begin{table*}
\resizebox{0.95\textwidth}{!}{\begin{minipage}{\textwidth}
\caption{X-ray mass estimates and comparison weak lensing masses.\label{tab-rmt500}}
\begin{tabular}{ c  r  l  r@{/}l  c  c  c  c }
\hline\\[-3.5ex]
\hline\\[-1.7ex]
		  & \multicolumn{1}{c}{Name}  	& \multicolumn{1}{c}{X-ray\_\_ID}  	&\multicolumn{2}{c}{$\chi^2$/ $d.o.f.$}	& $R_{500}$ 			 & $kT_{X,500}$ 				& $M_{500}$ 					& $M_{WL}$	\\
 		  & \multicolumn{1}{c}{} 	&  				&\multicolumn{2}{c}{}			& $^{\prime}\,$(kpc) 		 & keV  						& $10^{14}M_\odot$ 			& $10^{14}M_\odot$ 	\\[2pt]
\hline\\[-2ex]
\multirow{4}{*}[1.9em]{1.} 		  & \multirow{4}{*}[1.9em]{DLSCL~J0920.1+3029} 	 	& $^{1a}$\phm{I}CXOU~J092026+302938	& $1641$	& $1526$		& $4.11 \pm 0.05$($1103 \pm 15$) & $6.28^{+0.14}_{-0.14}$		&	$5.17^{+0.21}_{-0.21}$	& $3.39^{+0.18}_{-0.18}$	\\
		  &				& $^{1b}$\phm{I}CXOU~J092053+302800	& $1071$	& $964$			& $2.93 \pm 0.08$($~768 \pm 20$) & $3.02^{+0.16}_{-0.14}$		&	$1.72^{+0.14}_{-0.12}$	& $2.91^{+0.57}_{-0.38}$ 	\\
		  &				& $^{1c}$\phm{I}CXOU~J092110+302751	& $659$		& $494$			& $2.30 \pm 0.13$($~772 \pm 45$) & $3.61^{+0.39}_{-0.34}$		&	$2.09^{+0.36}_{-0.29}$	& $1.94^{+0.66}_{-0.57}$ 	\\
  		  &				& 
          \hspace{-2ex}$^{west}$XMMU~J091935+303155 		& $833$		& $768$			& $2.23 \pm 0.05$($~745\pm 16$)	 & $3.24^{+0.20}_{-0.19}$		&	$1.77^{+0.16}_{-0.16}$	& $1.8^{+1.0}_{-0.6}$$^{\left(a\right)}$  	\\
%----------------------------------------------------------------------------------------------------------------------------------------------------------------------------------------------------------------------------------------------------------
2. 		  & DLSCL~J0522.2$-$4820	& $^{2a}$\phm{I}CXOU~J052215$-$481816	& $144$		& $152$			& $3.34 \pm 0.09$($1020 \pm 23$) &  $4.03^{+0.37}_{-0.37}$	&	$2.67^{+0.39}_{-0.36}$	& $0.99^{+0.20}_{-0.39}$ 	\\ 
%---------------------------------------------------------------------------------------------------------------------------------------------------------------------------------------------------------------------------------------------------------
4. 		  & DLSCL~J1054.1$-$0549	& $^{\ \ \;}$\phm{I}CXOU~J105414$-$054849 	& $79$		& $63$			& $2.50 \pm 0.10$($~475 \pm 19$) & $1.05^{+0.06}_{-0.06}$		&	$0.36^{+0.04}_{-0.04}$	& $0.40^{+0.20}_{-0.20}$ 	\\
%----------------------------------------------------------------------------------------------------------------------------------------------------------------------------------------------------------------------------------------------------------
7. 		  & DLSCL~J0916.0+2931          & $^{7b}$\phm{I}CXOU~J091601+292750	& $74$		& $67$			& $1.43 \pm 0.08$($~540 \pm 32$) & $1.99^{+0.24}_{-0.19}$		&	$0.80^{+0.15}_{-0.12}$	& $0.10^{+0.30}_{-0.00}$ 	\\
%---------------------------------------------------------------------------------------------------------------------------------------------------------------------------------------------------------------------------------------------------------
8. 		  & DLSCL~J1055.2$-$0503	& $^{8b}$\phm{I}CXOU~J105535$-$045930	& $22$		& $17$			& $1.59 \pm 0.17$($~641 \pm 68$) & $3.08^{+0.67}_{-0.63}$		&	$1.47^{+0.51}_{-0.44}$	& $2.30^{+0.84}_{-0.84}$ 	\\
%---------------------------------------------------------------------------------------------------------------------------------------------------------------------------------------------------------------------------------------------------------
\multicolumn{9}{l}{\emph{Beyond the initial \citet{wittman06} publication}: }	\\[5pt]
$\;\;\;$B$9.\;\;$ & DLSCL~J1048.5$-$0411		& $^{\ \;\;}$XMMU~J104817$-$041233		& $25$		& $23$			& $3.13 \pm 0.31$($~728 \pm 72$) & $2.41^{+0.57}_{-0.42}$		&	$1.41^{+0.19}_{-0.46}$	& \nodata	\\
\hline
\end{tabular}
\end{minipage}
}
\tablecomments{
Superscripts to the left of the \xray\ IDs are identifiers given by \citetalias{abate09}, re-introduced here for easy reference within the mass comparison plot in Fig.~\ref{fig-mm}.
The $^{west}$ label refers to the Abell~781 ``west'' cluster, using the naming convention from \citet{sehg08}. 
The $^{\left(a\right)}$ indicates this mass is obtained from \citet{witt14} (see \S\ref{subsec-noic}). The X-ray--derived $M_{500}$ values use the \citet{vikh09} scaling relation. Uncertainties in weak lensing masses include both statistical and systematic effects (see \S~\ref{subsec-wl-masses}).}
\end{table*}%

Along with studying \xray\ properties ($L_X$, $T_X$, \& $M_X$) of weak lensing selected clusters, we do a direct comparison of mass estimates between weak lensing and \xray. 
For the weak lensing mass estimates, we take the  values obtained in \citet{abate09}.
Our \xray\ data are not of sufficient depth to estimate hydrostatic masses, however, they do allow \xray\ temperature to be used as a mass proxy for a fraction of the sample.
X-ray temperature correlates more tightly with the mass  \citep[e.g.,][]{vikh09}, and so we choose this over the luminosity as the proxy. 
We are unable to determine surface brightness profiles for most of the X-ray clusters so mass proxies such as the gas mass fraction, $f_g$, and the integrated gas mass times the temperature, $Y_X$, cannot be used. 

%%%%%%%%%%%%%%%%%%%%%%%%%%%%%%%%%%%%%%%%%%%%%%%%%%%%%%%%%%%%%%%%%%%%%%%
%		sub section - X-ray Masses
%%%%%%%%%%%%%%%%%%%%%%%%%%%%%%%%%%%%%%%%%%%%%%%%%%%%%%%%%%%%%%%%%%%%%%%

\subsection{X$-$ray Mass Estimates} \label{subsec-xmass}

We consider two $M_X-T_X$ relations to start, and proceed with two sets of \xray\ mass estimates. 
One relation is derived from \xmm\ data by \citet{arnaud2005}, and the other is derived from \chandra\ data by \citet{vikh09}.
Our aim here is to study any variation that may arise in our mass estimates from the choice of $M_X-T_X$ relation and to compare to published mass estimates when available.
The temperature measurement requires that the central core emission be removed, which means that we are able to get mass estimates for only nine X-ray clusters.

Both $M_X-T_X$ relations use masses within a fixed overdensity of $\Delta$=$500$ times the critical density, $\rho_{cr} = \frac{3 H^2}{8 \pi G}$.  
To obtain our own values of \rfive, we employ an iterative procedure.  
We start with a temperature from Table~\ref{tab-fit_res}, and estimate a mass from the $M_X-T_X$ relation.  
This mass then gives an \rfive\ via, $R_{\Delta} = \left( M_{tot}\left(T_X\right)/\left(\Delta \frac{4\pi}{3} \rho_{cr} \right) 
\right)^{1/3}$.
Using this \rfive\ we extract a new spectrum, excluding the core emission inside $r_i$. 
The inner and outer radii for extraction for the \citeauthor{arnaud2005} and \citeauthor{vikh09} relations are $r_i$$-$$r_o = 0.1R_{200}-0.5R_{200}$ and $r_i$$-$$r_o = 0.15R_{500}-R_{500}$ respectively.
The newly extracted spectrum gives a temperature inside the correct overdensity radius and produces the first estimates of \mfive\ and \rfive.  
Uncertainties in these are the result of propagating statistical measurement uncertainties of the spectral fit parameters as well as the uncertainties in the relation coefficients.
We repeat this process until subsequent \mfive\ and \rfive\  values converge to within measured uncertainty.
Data quality limits our ability to carry out this procedure for all clusters, and in Table~\ref{tab-rmt500} we report the masses (determined with the \citet{vikh09} relation) for the nine clusters  for which this was possible.

Both sets of \xray\ masses, from the two different scaling laws, agree well for the majority of the sample, but diverge at high temperatures, where several of the clusters in the Abell~$781$ complex lie.
To discriminate between the diverging estimates, we rely on the hydrostatic mass estimates obtained for Abell~$781$ by \citet{sehg08}.
The hydrostatic analysis is performed on combined \chandra\ and \xmm\ data, and so is not expected to favor either scaling law, a priori.  

We find that the hydrostatic masses agree with our estimates based on the \citet{vikh09} relation.  
Our measurement uncertainties (Table~\ref{tab-rmt500}) are lower than the \citet{sehg08} values, owing to the greater depth of the observation we analyze (Table~\ref{tab-obs_info}).
Since the rest of our clusters have agreeing mass estimates from the two scaling laws, we report only one set of \xray\ masses in Table~\ref{tab-rmt500}, determined with the \citet{vikh09} relation. 
Table~\ref{tab-rmt500} also lists the new core-excised temperatures along with \rfive\ and the goodness of fit indicators. 
Our choice to present masses from the \citet{vikh09} relation does not affect any conclusions we draw regarding the \xray\ and weak lensing mass comparison.
The resulting masses span an order of magnitude in range, with a median value of $\sim 2 \times 10^{14} M_\odot$.  

%%%%%%%%%%%%%%%%%%%%%%%%%%%%%%%%%%%%%%%%%%%%%%%%%%%%%%%%%%%%%%%%%%%%%%%
%		sub section - WL Masses
%%%%%%%%%%%%%%%%%%%%%%%%%%%%%%%%%%%%%%%%%%%%%%%%%%%%%%%%%%%%%%%%%%%%%%%

\subsection{Weak Lensing Mass Estimates} \label{subsec-wl-masses}

For all \xray\ counterparts of the DLS shear peaks published in \citeyear{wittman06}, \citet{abate09}, hereafter \citetalias{abate09}, obtained weak lensing mass estimates from the DLS data described in \citetalias{wittman06}.  
We briefly summarize their key steps here. 
For each \xray\ cluster, \citetalias{abate09} fit a mass distribution model based on the NFW mass density profile to its observed two dimensional shear profile.
Shear profiles were measured from source galaxy ellipticities, considering their full three dimensional positions (using photometric redshifts). 
Centers of these profiles were fixed to positions of the \xray\ peaks within a Gaussian window of 81 kpc (reported in table~1 of \citetalias{abate09}).    
The fits included uncertainties in shape noise, measurement noise, and photometric redshift noise and the final uncertainty on the mass has folded into it systematic uncertainties from effects such as biases in photometric redshift measurement or choice of mass profile center (within the 81 kpc window) that were estimated by modeling.

Where applicable, shear profiles were fit simultaneously for multiple neighboring \xray\ clumps, by adding shear linearly.  
The simultaneous fits account for the influence of neighboring mass concentrations on the shear of a given cluster and are thus believed to be more accurate than fitting each cluster individually.  
Their resulting masses are integrated out to an overdensity radius of $\Delta$=$200$ (table~3, \citetalias{abate09}), which we convert to masses within an overdensity radius of $\Delta$=$500$ assuming an NFW mass density profile and the observed mass concentration relations in \citet{duffy08}.
We list these masses in the last column of Table~\ref{tab-rmt500} for the seven clusters we use from \citetalias{abate09}.

We additionally include one weak lensing mass from \citet{witt14} because it was not in the \citetalias{abate09} study. 
\citet{witt14} perform a similar analysis fitting multiple shear profiles simultaneously, with centers guided by the X-ray peaks. The mass model is also based on the NFW profile, and their fitting incorporates a similar tomographic weighting to that in \citetalias{abate09}.

%%%%%%%%%%%%%%%%%%%%%%%%%%%%%%%%%%%%%%%%%%%%%%%%%%%%%%%%%%%%%%%%%%%%%%%
%		sub section - Mass Comp
%%%%%%%%%%%%%%%%%%%%%%%%%%%%%%%%%%%%%%%%%%%%%%%%%%%%%%%%%%%%%%%%%%%%%%%

\subsection{\xray\ $-$ Weak Lensing Mass Comparison} \label{subsec-mxm-comp}

We thus have a set of $8$ clusters with both \xray\ and weak lensing mass estimates for comparison. 
These eight clusters cover the full ranges of weak lensing masses, \xray\ temperatures (e.g., $T_X$ Table~\ref{tab-fit_res}), and redshifts of the full sample.

\begin{figure}
\includegraphics[width=\linewidth]{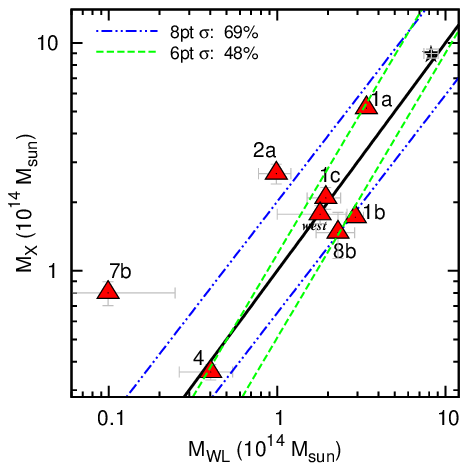}
\figcaption{%
The solid line shows equality. 
Intrinsic scatters determined including  (8pt) and excluding (6pt) the outliers (\S\ref{subsec-mxm-comp}) are plotted in the dash-dotted patterns; the corresponding best fit lines (not shown) are exactly in between the lines of scatter.
Labels refer to cluster IDs in Table~\ref{tab-rmt500}.
The star is the summed mass of Abell~781 (\S\ref{subsec-noic}). %
\label{fig-mm}}
\end{figure}

We plot the weak lensing and \xray\ \mfive\ values against each other in Figure~\ref{fig-mm}.  
The two sets of mass estimates are broadly consistent with each other, scattering on either side of equality.  
The scatter about equality is large and we identify two outliers based on the $M_X-M_{WL}$ fit.
We discuss the agreement, both overall and individually, between the weak lensing and \xray\ masses in detail below, beginning with some noteworthy cases. 

%%%%%%%%%%%%%%%%%%%%%%%%%%%%%%%%%%%
%     Figure Mx-Mwl
%%%%%%%%%%%%%%%%%%%%%%%%%%%%%%%%%%%

\subsubsection{Notes on Individual Comparisons}\label{subsec-noic}

The only shear peak in our sample (rank $4$) which has just one corresponding X-ray cluster, CXOU~J$105414-054849$, shows the best agreement in mass. 
The X-ray and shear-estimated masses for this cluster are in excellent agreement with $M^{WL}_{500}=0.40\pm0.2 \times 10^{14} M_\odot$ and $M^{X\textnormal{-}ray}_{500}=0.36\pm 0.04 \times 10^{14} M_\odot$.
The prominent case in our sample of shear resulting from superposed clusters (at different redshifts), Abell~$781$, presents with comparable masses when summed across the multiple components.
The three components with \citetalias{abate09} masses add to $\sum M^{WL}_{500} = 8.24 \pm 0.80 \times 10^{14} M_\odot$, with uncertainties crudely added in quadrature. 
The corresponding sum of X-ray masses is indeed crudely comparable at $\sum M^{X\textnormal{-}ray}_{500} = 8.98 \pm 0.41 \times 10^{14} M_\odot$ (see star point on mass comparison plot in Figure~\ref{fig-mm}).

The \xray\ mass we obtain for XMMU~J$091935$+$303155$, the ``West'' cluster in the Abell~$781$ complex, does not have a corresponding weak lensing mass in \citetalias{abate09}.
It does however have a weak lensing mass measurement in \citet{witt14}, and we use this value, $M^{WL}_{500} = 1.8^{+1.0}_{-0.6} \times 10^{14} M_\odot$, in the sample mass comparison section below. 
The mass comparison of this cluster, Abell~$781$~west, has been the subject of some controversy in the literature.
\citet{cook12} claimed that the weak lensing signal, based on three independent data sets including the DLS, was remarkably lower than expected based on the \citet{sehg08} \xray\ based mass estimate of  $M^{X\textnormal{-}ray}_{500}=2.2^{+0.5}_{-0.4} \times 10^{14} M_\odot$.  
\citet{witt14} then reviewed all available mass estimates including a DLS weak lensing estimate from \citet{sehg08}, a dynamical estimate from \citet{geller10}, and their own DLS weak lensing re-analysis and found that all estimates were consistent once uncertainties were properly treated.  
All estimates fell in the range $M_{500} = 0.8-2.2 \times 10^{14} M_\odot$, with the dynamical estimate at the low end and the X-ray estimate at the high end, but with no more than $2.2\sigma$ tension between them.
\citet{miyazaki15} found a weak lensing mass favoring the low end of this range, but still with uncertainties too large to rule out the higher values.
Our X-ray estimate here of $M^{X\textnormal{-}ray}_{500}=1.77\pm 0.16 \times 10^{14} M_\odot$  reduces the statistical uncertainties and places the mass in the middle to the upper half of the range seen in the literature.
We include this cluster in our mass comparison, using the weak lensing mass determined by \citet{witt14}.

There are outliers in the mass comparison plot of Figure~\ref{fig-mm}, which should carry a reduced weight in the sample mass comparison discussed below.
We demonstrate this here, by addressing them individually.
To start, the farthest outlier in the mass plot, marked 7b, does not have a well constrained weak lensing mass. 
On the other hand, its counterpart CXOU~J$091601+292750$, is well supported in the X-ray as a cluster, with a smooth surface distribution of photons, and a good fit to a spectrum at $z=0.53$ with $T_X = 2.09^{+.19}_{-.19}$ keV.  
Its luminosity and temperature are very close to the $L_X-T_X$ relation shown in Figure~\ref{fig-lt}.
In the weak lensing analysis, this cluster is not detected significantly; it is the farthest of three clusters associated with this shear peak and was likely picked up due to the  degree of smoothing of the shear field. 

The second outlier in Figure~\ref{fig-mm}, marked 2a, rests just outside the sample scatter, with a higher \xray\ mass.
This cluster, CXOUJ$052215$$-$$481816$, is a bright cluster in the \xray\ with robust measurements of luminosity and temperature.   
Our mass estimates depend on temperature as a proxy and are therefore subject to effects such as merger boosts. 
This cluster could be interacting with its neighbor, CXOUJ$052159$$-$$481606$, which could give it a boosted temperature, and result in an artificially higher \xray\ mass estimate.
Simulations \citep{rand02} show that mergers can affect both temperature and luminosity measurements such that this cluster may not appear as an outlier on the $L_X-T_X$ plot. 
And in fact, none of the clusters for which we compare masses are outliers on the $L_X-T_X$ relation in Figure~\ref{fig-lt} (marked with filled diamonds). 
Given the intrinsic limitations of determining masses in the two methods seen so far in the individual comparison, the X-ray to weak lensing mass comparison for the sample may not be low scatter.  
We quantify this in the next section.

\subsubsection{Overall Sample Comparison}

The eight clusters studied here show overall agreement between the \xray\ and weak lensing mass estimates, with considerable scatter.
We determine a linear relationship in log-space between the masses using the methods described in \citet[Eq. 35]{hogg10}.
We specifically choose this method, which allows us to estimate the intrinsic scatter of the data about the best fit relation,  in order to compare to the scatter determined for other cluster samples. 
The relationship we fit is of the form:
\begin{equation}
\log(M_X/10^{14}M_\odot) = a +  b\times\log(M_{WL}/10^{14}M_\odot).
\end{equation}
We convert our statistical uncertainties in mass to log-space and then symmetrize them, taking the average of absolute values.
We report parameters obtained both with and without the two statistical outliers in the sample, clusters  $2$a and $7$b.  

The best fit relation indicates that the X-ray and weak lensing masses are consistent with one another. 
The slope and intercept of our best fit relation for the case where we exclude outliers are $b = 1.26^{+0.53}_{-0.36}$ and  $a = -0.11^{+0.15}_{-0.21}$ which are consistent with the case of equality. 
Including the two outliers, the slope and intercept are $b = 0.96^{+0.48}_{-0.37}$  and $a = 0.05^{+0.12}_{-0.16}$. 
The intrinsic scatter of the points in the y-direction around this relation, is measured to be $48^{+39}_{-21}$\%, excluding the outliers. 
This scatter is plotted in Figure~\ref{fig-mm} (in dashed, green line). 
For comparison, we also plot the scatter determined from all eight points, $69^{+49}_{-25}$\%,  in Figure~\ref{fig-mm} (in dot-dashed, blue line)\footnote[7]{We also calculate scatter using the same $5$ \xray\ masses determined using the \citet{arnaud2005} $M_X-T_X$ law, and find a similar scatter of 53\%, which validates our earlier claim that the choice of $M_X-T_X$ law does not affect our conclusions.}.

We compare these results to the Canadian Cluster Comparison Project \citep[CCCP:][]{mahdavi13} who do a comparison of X-ray and weak lensing masses using $50$, massive, X-ray--selected clusters obtained from a large sky area.
Selection is the fundamental difference between the CCCP and our sample; this could result in systematic differences between the results. 
weak lensing--selection could be biased from, for example, line of sight mass projections, which could consistently boost the weak lensing masses we use here.
The CCCP sample contains many more massive clusters than our sample, with the high end of the CCCP range being twice as massive as our most massive cluster. 
Our mass range, however, excepting the lowest mass cluster, does fit comfortably within the CCCP range on its lower end.
Finally, the CCCP offer multiple mass estimates from X-ray and weak lensing, and we must select values determined compatibly to ours.

Although the published mass comparison by the CCCP is for masses measured inside an overdensity radius determined through weak lensing, they offer an online tool\footnote[8]{http://sfstar.sfsu.edu/cccp/, \citep{mahdavi13,hoekstra12} } attached to a database which allows us to make a comparison more consistently with what we do. 
Our X-ray masses are measured within an \rfive\ estimated with X-rays, and the weak lensing masses are measured within an overdensity radius estimated from weak lensing by profile fitting, which means our two mass estimates are independent.
The CCCP online tools offer access to weak lensing masses measured within a weak lensing estimated \rfive, and X-ray masses measured within an X-ray estimated \rfive, which are linked to the fitting algorithm \citep[based on][]{hogg10} that they have used in their paper.
We find that from all $50$ X-ray--selected CCCP clusters with masses measured like ours, the CCCP sample results in an intrinsic scatter of $58\% \pm 15\%$, which is fully consistent with the intrinsic scatter of the DLS shear-selected sample.  
This scatter is significantly larger than the value obtained from masses measured within identical radii:  $21\%\pm 6\%$ \citep{mahdavi13}; so the choice of overdensity radius is important in estimating the intrinsic scatter. 

The slope and intercept of our fitted mass relation is consistent with equality between X-ray and weak lensing masses, a trait that is also exhibited by the CCCP sample (although this sample shows mild, $\sim1\sigma$, indications of an X-ray underestimate).
The large uncertainties on our scaling law relation, however, mean that our mass-mass comparision is also compatible with a broad range of possible biases: for \xray\ hydrostatic bias see, e.g., \citet{vikh09}, \citet{mahdavi13}, \citet{donahue14}, or in the context of clusters selected via the Sunyaev-Zel'dovich effect, see, e.g., \citet{vonderlin14}, \citet{hoekstra15}, \citet{battaglia16}. 
Reducing the uncertainty on this comparison will require a much larger sample of shear-selected clusters, which will become available with future large area optical sky surveys, and targeted X-ray follow-up.

%%%%%%%%%%%%%%%%%%%%%%%%%%%%%%%%%%%%%%%%%%%%%%%%%%%%%%%%%%%%%%%%%%%%%%%
%%%%%%%%%%%%%%%%%%%%%%%%%%%%%%%%%%%%%%%%%%%%%%%%%%%%%%%%%%%%%%%%%%%%%%%
% %
% %		SECTION - SUMMARY
% %
%%%%%%%%%%%%%%%%%%%%%%%%%%%%%%%%%%%%%%%%%%%%%%%%%%%%%%%%%%%%%%%%%%%%%%%
%%%%%%%%%%%%%%%%%%%%%%%%%%%%%%%%%%%%%%%%%%%%%%%%%%%%%%%%%%%%%%%%%%%%%%%

\section{Summary}\label{sec-summ}

In this paper, we present the \xray\ properties and the weak lensing to \xray\ mass comparison of the first sample of shear-selected clusters \citep{wittman06}.
We report X-ray properties for 14 X-ray clusters that correspond to seven DLS shear peaks. 
An eighth DLS shear peak shows evidence for extended X-ray emission but the signal-to-noise for X-ray detection falls below our threshold for confirmation.
We additionally report properties of three X-ray clusters discovered in our fields which we cannot confidently associate to shear peaks. 

We determine luminosities and temperatures for $17$ X-ray clusters, and also determine a luminosity-temperature relation from $13$ of them with significant values of both $L_X$ and $T_X$ and that also correspond to the seven DLS shear peaks (Table~\ref{tab-fit_res}, Figure~\ref{fig-lt}). 
The clusters have widely varying \xray\ properties; a factor of 6 in temperature and four orders of magnitude in luminosity. 
The ranges of redshift and mass of the sample are also substantial (\S\ref{subsec-prop}). 
The best fit $L_X-T_X$ relation is consistent with X-ray cluster samples selected without regard for dynamical state as well as with the weak lensing selected sample of \citet{giles15}.  
Unlike this other weak lensing study, however, we find that the DLS X-ray clusters are inconsistent with a self-similar slope for the  $L_X-T_X$ relation (our slope is $2.93 \pm 0.15$).

We determine \xray\ mass estimates using the \citet{vikh09} \xray\ mass-temperature relation.  
Core-excluded temperatures required for this estimate can be constrained for nine of our clusters. 
Weak lensing mass estimates are available for eight of them, with seven determined by \citet{abate09} by fitting mass profiles centered at the X-ray peaks in \chandra\ data (analyzed in \citetalias{wittman06}).
An eighth weak lensing mass is available from \citet{witt14} which we include in our mass comparison.
We find overall agreement between the \xray\ and weak lensing masses.
The sample is characterized by an intrinsic scatter of $\sim 48\%$ with large uncertainty about the best fit mass relation; this is consistent with the \citet{mahdavi13} X-ray selected sample whose mass range largely overlaps with our sample.  

We summarize some of the issues related to shear selection based on this study and other earlier work on the DLS. 
A major difference with other selection techniques is the association of multiple X-ray clusters with a single weak lensing shear peak --- this complicates the identification of X-ray with shear.
We find the shear associated X-ray clusters are not necessarily high mass individuals, and in fact, they cover an order of magnitude range in mass.
Our $L_X-T_X$ relation is consistent with other X-ray cluster samples selected without regard to their dynamical activity, but is inconsistent with the self-similar relation.
Weak lensing and X-ray masses determined individually for each shear--associated X-ray cluster agree broadly, and exhibit intrinsic scatter that is consistent with X-ray selected samples, as long as the two mass estimates are determined independently from one another.

Currently the number of individual, well studied, X-ray clusters from weak lensing selected samples is small, which is a consequence of the lack of large area, deep optical weak lensing surveys.
As we approach the era of the Large Synoptic Survey Telescope, this issue will be alleviated.

%%%%%%%%%%%%%%%%%%%%%%%%%%%%%%%%%%%%%%%%%%%%%%%%%%%%%%%%%%%%%%%%%%%%%%%
%%%%%%%%%%%%%%%%%%%%%%%%%%%%%%%%%%%%%%%%%%%%%%%%%%%%%%%%%%%%%%%%%%%%%%%
% %
% %		ACKNOWLEDGEMENTS 
% %
%%%%%%%%%%%%%%%%%%%%%%%%%%%%%%%%%%%%%%%%%%%%%%%%%%%%%%%%%%%%%%%%%%%%%%%
%%%%%%%%%%%%%%%%%%%%%%%%%%%%%%%%%%%%%%%%%%%%%%%%%%%%%%%%%%%%%%%%%%%%%%%

\acknowledgments
This work was partially supported by NASA grants NNG05GR44G, NNX07AU67G and NNX09AP40G.  We acknowledge Tony Tyson, Ian Dell'Antonio, David Spergel, Joe Hennawi, Vera Margoniner, Gillian Wilson, Dara Norman, and Neelima Sehgal for help with the original \xmm\ proposals. AJD would like to acknowledge Amitpal Tagore for helpful discussions.

%%%%%%%%%%%%%%%%%%%%%%%%%%%%%%%%%%%%%%%%%%%%%%%%%%%%%%%%%%%%%%%%%%%%%%%
%%%%%%%%%%%%%%%%%%%%%%%%%%%%%%%%%%%%%%%%%%%%%%%%%%%%%%%%%%%%%%%%%%%%%%%
% %
% %		BIBLIOGRAPHY 
% %
%%%%%%%%%%%%%%%%%%%%%%%%%%%%%%%%%%%%%%%%%%%%%%%%%%%%%%%%%%%%%%%%%%%%%%%
%%%%%%%%%%%%%%%%%%%%%%%%%%%%%%%%%%%%%%%%%%%%%%%%%%%%%%%%%%%%%%%%%%%%%%%

\end{document}